\documentclass[aps,preprint,amsmath,amssymb,floatfix,showpacs,
superscriptaddress]{revtex4}

\usepackage[dvips]{graphics}
\usepackage{color}
\definecolor{gold}{rgb}{0.85,0.66,0}
\definecolor{dred}{rgb}{0.7,0,0}

\begin{document}

\title{\textcolor{dred}{Mesoscopic Physics and Nanoelectronics}}

\author{Santanu K. Maiti}

\email{santanu.maiti@saha.ac.in}

\affiliation{Theoretical Condensed Matter Physics Division, Saha
Institute of Nuclear Physics, Sector-I, Block-AF, Bidhannagar,
Kolkata-700 064, India}

\affiliation{Department of Physics, Narasinha Dutt College, 129
Belilious Road, Howrah-711 101, India}

\begin{abstract}
Electronic transport properties through some model quantum systems 
are re-visited. A simple tight-binding framework is given to describe
the systems where all numerical calculations are made using the Green's 
function formalism. First, we demonstrate electronic transport in four 
different polycyclic hydrocarbon molecules, namely, benzene, napthalene, 
anthracene and tetracene. It is observed that electron conduction through 
these molecular wires is highly sensitive to molecule-to-electrode 
coupling strength and quantum interference of electronic waves passing 
through different branches of the molecular ring. Our investigations 
predict that to design a molecular electronic device, in addition to 
the molecule itself, both the molecular coupling and molecule-to-electrode 
interface geometry are highly important. Next, we make an in-depth study 
to design classical logic gates with the help of simple mesoscopic rings, 
based on the concept of Aharonov-Bohm effect. A single mesoscopic ring or 
two such rings are used to establish the logical operations where the key 
controlling parameter is the magnetic flux threaded by the ring. The 
analysis might be helpful in fabricating meso-scale or nano-scale logic 
gates. Finally, we address multi-terminal quantum transport through a single
benzene molecule using Landauer-B\"{u}ttiker formalism. Quite interestingly
we see that a three-terminal benzene molecule can be operated as an
electronic transistor and this phenomenon is justified through 
current-voltage characteristics. All these essential features of electron
transport may provide a basic theoretical framework to examine electron
conduction through any multi-terminal quantum system.
\end{abstract}


\maketitle

\section{Introduction}

\subsection{\textcolor{dred}{Basic concepts}}

Mesoscopic physics is a sub-discipline of condensed matter physics which
deals with systems whose dimensions are intermediate between the
microscopic and macroscopic length scales~\cite{imrybook,datta1,datta2,
metalidis,weinmann}. In meso-scale region fluctuations play an important 
role and the systems are treated quantum mechanically, in contrast to 
the macroscopic objects where usually the laws of classical mechanics 
are used. In a simple version we can say that a macroscopic system 
when scaled down to a meso-scale starts exhibiting quantum mechanical 
phenomena. The most relevant length scale of quantifying a mesoscopic 
system is probably the phase coherence length $L_{\phi}$, the length 
scale over which the carriers preserve their phase information. This 
phase coherence length is, on the other hand, highly sensitive to 
temperature and sharply decreases with the rise of temperature. 
Therefore, to be in the mesoscopic regime we have to lower the 
temperature sufficiently (of the order of liquid He) such that phase 
randomization process caused by phonons gets minimum. Though there is 
no such proper definition of the mesoscopic region, but the studied 
mesoscopic objects are normally in the range of $100$-$1000$ nanometers.

Several spectacular effects appear as a consequence of quantum phase 
coherence of the electronic wave functions in mesoscopic systems like
one-dimensional ($1$D) quantum wires, quantum dots where electrons are 
fully confined, two-dimensional ($2$D) electron gases in heterostructures, 
etc. For our illustrative purposes here we describe very briefly some 
of these issues.

\vskip 0.2cm
\noindent
(i) {\bf \textcolor{dred}{Aharonov-Bohm Effect:}} One of the most 
significant experiments in mesoscopic physics is the observation of 
Aharonov-Bohm (AB) oscillations in conductance of a small metallic 
ring threaded by a magnetic flux $\phi$~\cite{webb1,webb2}. The 
origin of conductance oscillations lies in the quantum interference 
among the waves traversing through two arms of the ring. This 
pioneering experiment has opened up a wide range of challenging 
and new physical concepts in the mesoscopic regime.

\vskip 0.2cm
\noindent
(ii) {\bf \textcolor{dred}{Conductance Fluctuations:}} The pronounced 
fluctuations in 
conductance of a disordered system are observed when the temperature is 
lowered below $1$K~\cite{webb3}. These fluctuations are originated from 
the interference effects of the electronic wave functions traveling across 
the system and are fully different from the fluctuations observed in 
traditional macroscopically large objects. The notable feature of 
conductance fluctuations in the mesoscopic regime is that their 
magnitudes are always of the order of the conductance quantum $e^2/h$,
and accordingly, these fluctuations are treated as `universal conductance
fluctuations'~\cite{fuku}.  

\vskip 0.2cm
\noindent
(iii) {\bf \textcolor{dred}{Persistent Current:}} In thermodynamic 
equilibrium a 
small metallic ring threaded by magnetic flux $\phi$ supports a current 
that does not decay dissipatively even at non-zero temperature. It is 
the well-known phenomenon of persistent current in mesoscopic normal 
metal rings~\cite{butt,levy,mailly1,chand,jari,deb,butt1,cheu1,cheu2,blu,
san1,san2,san3,san4,san5}. This is a purely quantum mechanical effect and 
gives an obvious demonstration of the AB effect~\cite{aharo}.

\vskip 0.2cm
\noindent
(iv) {\bf \textcolor{dred}{Integer Quantum Hall Effect:}} The integer 
quantum Hall effect is probably the best example of quantum phase 
coherence of electronic wave functions in two-dimensional electron gas 
(2DEG) systems~\cite{klit}. In the Hall experiment, a current is 
allowed to pass through a conductor (2DEG), and, 
the longitudinal voltage $V_x$ and transverse Hall voltage $V_H$ are
measured as a function of the applied magnetic field $B$ which is 
perpendicular to the plane of the conductor. In the limit of weak magnetic
field, the Hall resistance $R_H$ varies linearly with the field strength
$B$, while the longitudinal resistance $R_x$ remains unaffected by this
field. These features can be explained by the classical Drude model.
On the other hand for strong magnetic field and in the limit of low
temperature a completely different behavior is observed and the classical
Drude model fails to explain the results. In high magnetic field $R_x$ 
shows oscillatory nature, while $R_H$ shows step-like behavior with sharp 
plateaus. On these plateaus the values of $R_H$ are given by $h/n e^2$,
$n$ being an integer, and they are highly reproducible with great precision.
These values are extremely robust so that they are often used as the 
standard of resistance. The integer quantum Hall effect is a purely 
quantum mechanical phenomenon due to the formation of the Landau levels
and many good reviews on IQHE are available in the literature~\cite{imrybook,
pran,chakra}.

\vskip 0.2cm
\noindent
(v) {\bf \textcolor{dred}{Fractional Quantum Hall Effect:}} Unlike the 
integer quantum Hall 
effect, at too high magnetic fields and low temperatures, a two-dimensional
electron gas shows additional plateaus in the Hall resistance at fractional
filling factors~\cite{tsui}. It has been verified that the Coulomb 
correlation between the electrons becomes important for the interpretation 
of the fractional quantum Hall effect and the presence of fractional 
filling has been traced back to the existence of correlated collective 
quasi-particle excitations~\cite{laug}. Extensive reviews on this topic 
can be found in the literature~\cite{chakra}.

In the mesoscopic regime, electronic transport cannot be investigated by
using the conventional Boltzmann transport equation since at this length
scale quantum phase coherence plays an important role and a full quantum
mechanical treatment is needed~\cite{datta1,datta2}. In an `open system' 
the Landauer approach~\cite{land}
of `two-terminal conductance' provides an elegant technique to reveal the
transport mechanisms. In the Landauer formalism a quantum system is 
sandwiched between two macroscopic reservoirs, the so-called electrodes, 
those are kept at thermal equilibrium. By applying a bias voltage we tune 
the chemical potentials of these electrodes. The main signature of the
electrodes is that electrons passing through them along the 
longitudinal direction can be described as plane waves and suffer no
backscattering whatsoever. This allows us to describe the properties of the
conductor in terms of its scattering matrix $S$ on the basis of plane waves
within the electrodes. The Landauer formulation is probably the simplest
and elegant approach for studying electron transport in low-dimensional
quantum systems. 

Ongoing trend of miniaturizing electronic devices eventually approaches the 
ultimate limit where even a single molecule can be used as an electrical 
circuit element. Idea of devicing a single molecule as the building block 
of future generation electronics seems fascinating because of the 
possibility to assemble a large number of molecules onto a chip i.e., 
remarkable enhancement in integration density can take place~\cite{ratner1}. 
Discovery of sophisticated molecular scale measurement 
methodologies such as scanning tunneling microscopy (STM), atomic force 
microscopy (AFM), scanning electro-chemical microscopy (SECM), etc., 
have made it possible to study electron transport phenomena in molecular 
bridge systems~\cite{chen}.

The idea of using molecules as active components of a device was suggested
by Aviram and Ratner~\cite{aviram} over three decades ago. Since then 
several ab-initio and model calculations have been performed to 
investigate molecular transport theoretically~\cite{ventra1,ventra2,sumit,
tagami,orella1,orella2,arai,baer1,baer2,baer3,walc1,walc2,tagami1,tagami2,
san6,san7,san8,san9,san10,san11,noza,maas}. But experimental 
realizations took a little longer time to get feasible. In $1997$, Reed 
and co-workers~\cite{reed1} have studied the current-voltage ($I$-$V$) 
characteristics of a single benzene molecule attached to electrodes 
via thiol groups. Later various other experiments have been made to explore 
many interesting features e.g., ballistic transport, quantized conductance, 
negative differential resistance (NDR)~\cite{saiful}, molecular
transistor operation~\cite{park,tao} to name a few.

In short we can say that the rapid progress of theoretical as well as 
experimental works on mesoscopic physics over the last few decades proves 
that it is a highly exciting and challenging branch of condensed matter 
physics and we hope that it will be continuing for many more decades.

\subsection{\textcolor{dred}{Aim of the review}}

In this dissertation we address several important issues on electron 
transport through some meso-scale systems which are quite challenging 
from the standpoint of theoretical as well as experimental research. 
A brief outline of the presentation is as follow.

On the meso-scale organic molecules, cluster of atoms, quantum dots, carbon
nanotubes, etc., can be produced with flexible and tunable conduction
properties and these new realities have tremendous technological importance.
The physics of electron transport through such devices is surprisingly rich.
Many fundamental experimentally observed phenomena in such devices can be 
understood by using simple arguments. In particular, the formal relation 
between conductance and transmission coefficients (the Landauer formula) 
has enhanced the understanding of electronic transport in the molecular 
bridge system. To reveal these facts, in the first part of this review 
we investigate electron transport properties of some molecular bridge 
systems within the tight-binding framework using Green's function technique 
and try to explain the behavior of electron conduction in the aspects of 
quantum interference of electronic wave functions, molecule-to-electrode 
coupling strength, molecular length, etc. Our model calculations provide 
a physical insight to the behavior of electron conduction through molecular 
bridge systems.

Next, we explore the possibilities of designing classical logic gates at 
meso-scale level using simple mesoscopic rings. A single 
ring is used for designing OR, NOT, XOR, XNOR and NAND gates, while AND 
and NOR gate responses are achieved using two such rings and in all these 
cases each ring is threaded by a magnetic flux $\phi$ which plays the 
central role in the logic gate operation. We adopt a simple tight-binding 
Hamiltonian to describe the model where a mesoscopic ring is attached to 
two semi-infinite one-dimensional non-magnetic electrodes. Based on single
particle Green's function formalism all calculations which describe
two-terminal conductance and current through the quantum ring are
performed numerically. The analysis may be helpful in fabricating
mesoscopic or nano-scale logic gates.

Finally, in the last part, we focus our attention on the multi-terminal 
transport problem. Here we no longer use the Landauer approach. B\"{u}ttiker 
has extended the study of two-terminal quantum transport into the 
multi-terminal case which is known as Landauer-B\"{u}ttiker formalism. 
With this approach we study multi-terminal quantum transport in a 
single benzene molecule to examine the transport properties in terms of
conductance, reflection probability and current-voltage characteristics.

\section{Two-terminal molecular transport}

Molecular electronics is an essential technological concept of fast-growing 
interest since molecules constitute promising building blocks for future 
generation of electronic devices where electronic transport becomes 
coherent~\cite{ratner1,nitzan1}. For purposeful design of an electronic
circuit using a single molecule or a cluster of molecules, the most 
important requirement is the understanding of fundamental processes of 
electron conduction through separate molecules used in the circuit.
A fruitful discussion of electron transport in a molecular wire was first
studied theoretically by Aviram and Ratner during 1974~\cite{aviram}.
Later, numerous experiments have been made in different molecules placed
between electrodes with few nano-meter separation~\cite{reed1,saiful,park,
tao,metz,fish,reed2,tali,sch,mir1,mir2,maji,weiss}. It is very crucial
to control electron transmission through such molecular electronic devices,
and, though extensive studies have been done, yet the present understanding
about it is not fully explored. For example, it is not very clear how
molecular conduction is affected by geometry of the molecule itself or
by the nature of its coupling to side attached electrodes. To construct 
an electronic device made with molecules and to utilize it properly we need
a deep analysis of structure-conductance relationship. In a recent work
Ernzerhof {\em et al.}~\cite{ern1} have illustrated a general design 
principle based on several model calculations to reveal this concept.
In presence of applied bias voltage a current passes through the 
molecule-electrode junction and it becomes a non-linear function of
the applied voltage. The detailed description of it is quite complicated.
Electron transport properties in molecular systems are highly sensitive
on several quantum effects like quantum interference of electronic waves
passing through different arms of the molecular rings, quantization of
energy levels, etc.~\cite{baer1,baer2,baer3,walc1,walc2}. The main 
motivation of studying molecular transport 
is that molecules are currently the subject of substantial theoretical,
experimental and technological interest. Using molecules we can design 
logic gates, molecular switches, several transport elements that need to
be well characterized and explained.

In this section we address the behavior of electron transport through some 
polycyclic hydrocarbon molecules based on a simple tight-binding framework.
Though several {\em ab initio}~\cite{yal,ven,xue,tay,der,dam,ern3,zhu1,zhu2} 
methods are used to enumerate electron transport in molecular systems, but 
simple parametric approaches are rather much helpful to understand the basic 
mechanisms of electron transport in detail. Beside this here we also make
our attention only on the qualitative effects instead of the quantitative
results which also motivate us to perform model calculations in the transport
problem.

To describe the behavior of molecular transport let us first construct 
the methodology for two-terminal quantum transport through a simple
finite sized conductor.

\subsection{\textcolor{dred}{Theoretical Formulation}}

\begin{figure}[ht]
{\centering \resizebox*{15cm}{3.5cm}{\includegraphics{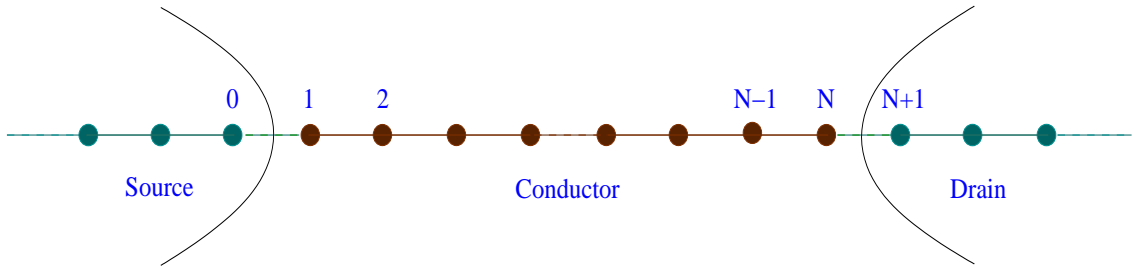}}\par}
\caption{Schematic view of a one-dimensional conductor with $N$ number
of atomic sites attached to two electrodes, namely, source and drain. 
The atomic sites at the two extreme ends of the conductor are labeled 
as $1$ and $N$, respectively.}
\label{wire}
\end{figure}

We begin with Fig.~\ref{wire}. A finite sized $1$D conductor with $N$ 
atomic sites is attached to two semi-infinite $1$D metallic electrodes, 
viz, source and drain. At much low temperature and bias voltage, 
conductance $g$ of the conductor can be written by using Landauer 
conductance formula,
\begin{equation}
g=\frac{2e^2}{h}T
\label{land}
\end{equation}
where, $T$ is the transmission probability of an electron through the
conductor. In terms of the Green's function of the conductor and its 
coupling to the side attached electrodes, transmission probability
can be expressed as,
\begin{equation}
T={\mbox{Tr}} \left[\Gamma_S G_c^r \Gamma_D G_c^a\right]
\label{trans1}
\end{equation}
where, $G_c^r$ and $G_c^a$ are the retarded and advanced Green's functions
of the conductor, respectively. $\Gamma_S$ and $\Gamma_D$ are the coupling
matrices due to the coupling of the conductor to the source and drain,
respectively. For the combined system i.e., the conductor and two 
electrodes, the Green's function becomes,
\begin{equation}
G=\left(E-H\right)^{-1}.
\end{equation}
$E$ is the injecting energy of the source electron and $H$ is the 
Hamiltonian of the combined system. Evaluation of this Green's function 
requires the inversion of an infinite matrix as the system consists of 
the finite size conductor and two semi-infinite $1$D electrodes, which 
is really a very difficult task. However, the entire system can be 
partitioned into sub-matrices corresponding to the individual sub-systems, 
and then the Green's function for the conductor can be effectively 
written as,
\begin{equation}
G_c=\left(E-H_c-\Sigma_S-\Sigma_D\right)^{-1}
\label{grc}
\end{equation}
where, $H_c$ is the Hamiltonian of the conductor. Withing a 
non-interacting picture, the tight-binding Hamiltonian of the conductor 
looks like,
\begin{equation}
H_c=\sum_i \epsilon_i c_i^{\dagger} c_i + \sum_{<ij>}t
\left(c_i^{\dagger}c_j + c_j^{\dagger}c_i \right).
\label{hamil1}
\end{equation}
$c_i^{\dagger}$ ($c_i$) is the creation (annihilation) operator of an 
electron at site $i$, $\epsilon_i$ is the site energy of an electron 
at the $i$-th site and $t$ corresponds to the nearest-neighbor hopping 
integral. A similar kind of tight-binding Hamiltonian is also used for
the description of electrodes where the Hamiltonian is parametrized by
constant on-site potential $\epsilon_0$ and nearest-neighbor hopping
integral $v$. In Eq.~\ref{grc},
$\Sigma_S=h_{Sc}^{\dagger} g_S h_{Sc}$ and $\Sigma_D=h_{Dc} g_D
h_{Dc}^{\dagger}$ are the self-energy operators due to the two
electrodes, where $g_S$ and $g_D$ are the Green's functions for the
source and drain, respectively. $h_{Sc}$ and $h_{Dc}$ are the coupling
matrices and they will be non-zero only for the adjacent points in the
conductor, $1$ and $N$ as shown in Fig.~\ref{wire}, and the electrodes
respectively. The coupling terms $\Gamma_S$ and $\Gamma_D$ of the
conductor can be calculated from the following expression,
\begin{equation}
\Gamma_{\{S,D\}}=i\left[\Sigma_{\{S,D\}}^r-\Sigma_{\{S,D\}}^a\right].
\end{equation}
Here, $\Sigma_{\{S,D\}}^r$ and $\Sigma_{\{S,D\}}^a$ are the retarded and
advanced self-energies, respectively, and they are conjugate to each
other. Datta {\em et al.}~\cite{tian} have shown that the self-energies
can be expressed like,
\begin{equation}
\Sigma_{\{S,D\}}^r=\Lambda_{\{S,D\}}-i \Delta_{\{S,D\}}
\end{equation}
where, $\Lambda_{\{S,D\}}$ are the real parts of the self-energies which
correspond to the shift of the energy eigenvalues of the conductor and the
imaginary parts $\Delta_{\{S,D\}}$ of the self-energies represent the
broadening of the energy levels. Since this broadening is much larger
than the thermal broadening we restrict our all calculations only at
absolute zero temperature. The real and imaginary parts of the
self-energies can be determined in terms of the hopping integral
($\tau_{\{S,D\}}$) between the boundary sites ($1$ and $N$) of the
conductor and electrodes, energy ($E$) of the transmitting electron and
hopping strength ($v$) between nearest-neighbor sites of the electrodes.

The coupling terms $\Gamma_S$ and $\Gamma_D$ can be written in terms of
the retarded self-energy as,
\begin{equation}
\Gamma_{\{S,D\}}=-2\, {\mbox{Im}} \left[\Sigma_{\{S,D\}}^r\right].
\end{equation}
Now all the information regarding the conductor to electrode coupling
are included into these two self energies as stated above. Thus, by 
calculating the self-energies, the coupling terms $\Gamma_S$ and $\Gamma_D$ 
can be easily obtained and then the transmission probability ($T$) will be 
calculated from the expression as presented in Eq.~\ref{trans1}.

Since the coupling matrices $h_{Sc}$ and $h_{Dc}$ are non-zero only for the
adjacent points in the conductor, $1$ and $N$ as shown in Fig.~\ref{wire},
the transmission probability becomes,
\begin{equation}
T(E)=4\,\Delta_{11}^S(E)\, \Delta_{NN}^D(E)\,|G_{1N}(E)|^2
\label{trans2}
\end{equation}
where, $\Delta_{11}=<1|\Delta|1>$, $\Delta_{NN}=<N|\Delta|N>$ and
$G_{1N}=<1|G_c|N>$.

The current passing through the conductor is treated as a single-electron
scattering process between the two reservoirs of charge carriers. We
establish the current-voltage relation from the 
expression~\cite{datta1,datta2},
\begin{equation}
I(V)=\frac{e}{\pi \hbar}\int \limits_{E_F-eV/2}^{E_F+eV/2} T(E)\,dE
\end{equation}
where, $E_F$ is the equilibrium Fermi energy. For the sake of simplicity,
here we assume that the entire voltage is dropped across the
conductor-electrode interfaces and it doesn't greatly affect the
qualitative aspects of the current-voltage characteristics. This is
due to the fact that the electric field inside the conductor,
especially for shorter conductors, seems to have a minimal effect on
the conductance-voltage characteristics. On the other hand, for quite
larger conductors and higher bias voltages, the electric field inside
the conductor may play a more significant role depending on the internal
structure of the conductor~\cite{tian}, though the effect becomes too
small. Using the expression of $T(E)$ (Eq.~\ref{trans2}), the final form
of $I(V)$ is,
\begin{eqnarray}
I(V) = \frac{4e}{\pi \hbar}\int \limits_{E_F-eV/2}^{E_F+eV/2}
\Delta_{11}^S(E)\, \Delta_{NN}^D(E) \times \, |G_{1N}(E)|^2 \,dE
\label{curr}
\end{eqnarray}
Eqs.~\ref{land}, \ref{trans2} and \ref{curr} are the final working
expressions for the determination of conductance $g$, transmission 
probability $T$, and current $I$, respectively, through any finite 
sized conductor placed between two $1$D metallic reservoirs.
Throughout our presentation, we use the units where $c=h=e=1$, and, the
energy scale is measured in unit of $t$.

\subsection{\textcolor{dred}{Molecular System and Transport Properties}}

\begin{figure}[ht]
{\centering \resizebox*{8.5cm}{8.5cm}{\includegraphics{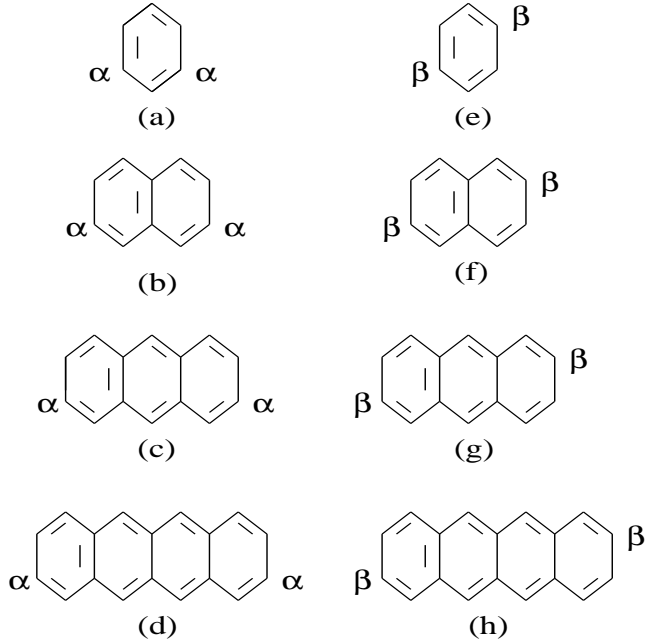}}\par}
\caption{Schematic view of four different polycyclic hydrocarbon molecules: 
benzene, napthalene, anthracene and tetracene. The molecules are connected 
to the electrodes in two different configurations via thiol (S-H bond) 
groups. One is the so-called {\em cis} configuration ($\alpha$-$\alpha$ 
position) and the other one is the so-called {\em trans} configuration 
($\beta$-$\beta$ position).}
\label{hydro}
\end{figure}

Based on the above two-terminal transport formulation, in this section, 
we describe the behavior of electron conduction through some polycyclic 
hydrocarbon molecules those are schematically shown in Fig.~\ref{hydro}.
The molecules are: benzene (one ring), napthalene (two rings), anthracene 
(three rings) and tetracene (four rings). These molecules are connected to
the electrodes (source and drain) via thiol (S-H bond) groups. In
real experimental situations, gold (Au) electrodes are generally used and
the molecules are coupled to them through thiol groups in the chemisorption
technique where hydrogen (H) atoms removes ans sulfur (S) atoms reside.
To emphasize the effect of quantum interference on electron transport,
we couple the molecules to the source and drain in two different 
configurations. One is defined as {\em cis} configuration where two
electrodes are placed at the $\alpha$ sites, whereas in the other arrangement, 
called as {\em trans} configuration, electrodes are connected at the $\beta$ 
sites. Molecular coupling is another important factor which controls the
electron transport. To justify this fact, here we describe the essential
features of electron transport for the two limiting cases of molecular
coupling. One is the weak-coupling limit which mathematically treated as
$\tau_{\{S,D\}} << t$ and in the other case we have $\tau_{\{S,D\}} 
\sim t$ which is defined as the strong-coupling limit. $\tau_S$ and 
$\tau_D$ are the hopping strengths of the molecule to the source and 
drain, respectively. The common set of values of the parameters used 
in the calculation are as follows. $\tau_S=\tau_D=0.5$, $t=2.5$ 
(weak-coupling) and $\tau_S=\tau_D=2$, $t=2.5$ (strong-coupling). In 
the electrodes we set $\epsilon_0=0$ and $v=4$. The Fermi energy $E_F$ 
is fixed at $0$.

\subsubsection{\textcolor{dred}{Conductance-energy characteristics}}

\begin{figure}[ht]
{\centering \resizebox*{11cm}{9cm}{\includegraphics{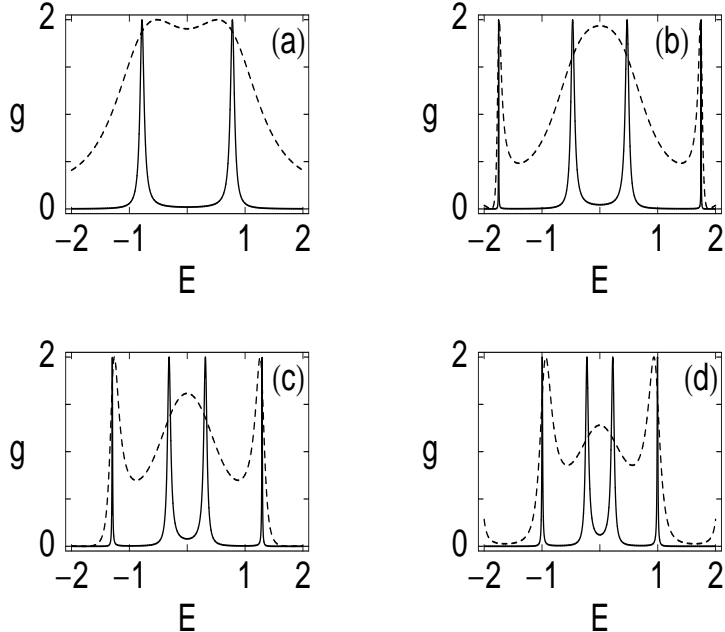}}\par}
\caption{$g$-$E$ spectra of four different polycyclic hydrocarbon 
molecules attached to the electrodes in the {\em trans} configuration, 
where (a), (b), (c) and (d) correspond to the benzene, napthalene, 
anthracene and tetracene molecules, respectively. The solid and dotted 
curves represent the weak and strong molecule-to-electrode coupling limits, 
respectively.}
\label{transcond}
\end{figure}

In Fig.~\ref{transcond}, we show the behavior of conductance ($g$) as a 
function of injecting electron energy ($E$) for the hydrocarbon molecules 
when they are coupled to the electrodes in the {\em trans} configuration,
where (a), (b), (c) and (d) correspond to the benzene, napthalene, 
anthracene and tetracene molecules, respectively. In the limit of weak 
molecular coupling, conductance shows
sharp resonant peaks (solid lines) for some specific energy eigenvalues,
while it drops almost to zero for all other energies. At the resonance, 
conductance approaches to $2$, and therefore, the transmission probability 
($T$) becomes unity (from the Landauer conductance formula $g=2T$ in our
chosen unit system $c=e=h=1$). The resonant peaks in the
conductance spectrum coincide with eigenenergies of the single
hydrocarbon molecules. Thus, from the conductance spectrum we can easily
implement the electronic structure of a molecule. The nature of these 
resonant peaks gets significantly modified when molecular coupling is
increased. In the strong-coupling limit, width of the resonant peaks 
gets broadened as shown by the dotted curves in Fig.~\ref{transcond} and it 
emphasizes that electron conduction takes place almost for all energy 
values. This is due to the broadening of the molecular energy levels, 
where contribution comes from the imaginary parts of the self-energies 
$\Sigma_{S(D)}$~\cite{tian}.

\begin{figure}[ht]
{\centering \resizebox*{11cm}{9cm}{\includegraphics{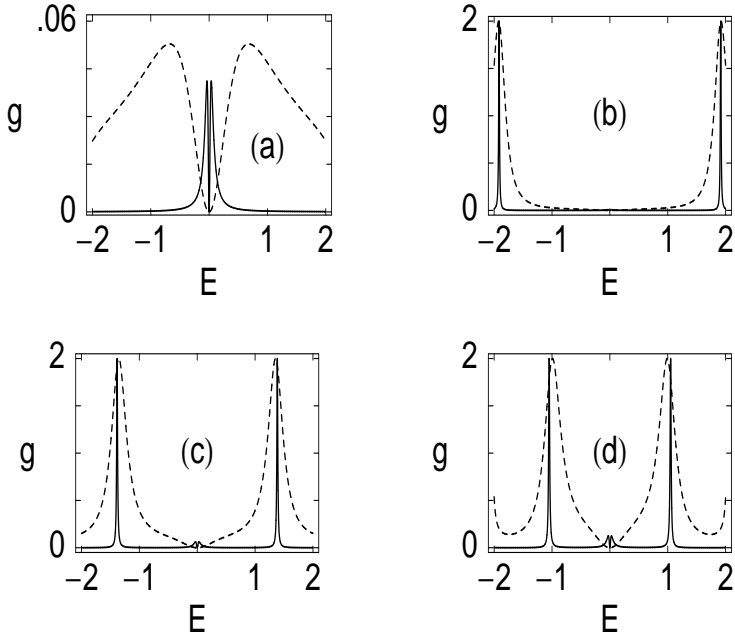}}\par}
\caption{$g$-$E$ curves of four different polycyclic hydrocarbon 
molecules attached to the electrodes in the {\em cis} configuration,
where (a), (b), (c) and (d) correspond to the benzene, napthalene,
anthracene and tetracene molecules, respectively. The solid and dotted
lines represent the similar meaning as in Fig.~\ref{transcond}.}
\label{ciscond}
\end{figure}

To establish the effect of quantum interference on electron transport,
in Fig.~\ref{ciscond}, we plot conductance-energy characteristics for
these molecules when they are connected to the electrodes in the {\em cis} 
configuration. (a), (b), (c) and (d) correspond to the results for the 
benzene, napthalene, anthracene and tetracene molecules, respectively, 
where the solid and dotted curves indicate the identical meaning as in 
Fig.~\ref{transcond}.
From these spectra we clearly see that some of the conductance peaks do 
not reach to unity anymore and achieve much reduced amplitude. This 
behavior can be justified as follow. During the propagation of electrons
from the source to drain, electronic waves which pass through different
arms of the molecular ring/rings can suffer a phase shift among themselves,
according to the result of quantum interference. As a result, probability
amplitude of getting an electron across the molecule becomes increased or
decreased. The cancellation of transmission probabilities emphasizes
the appearance of anti-resonant states which provides an interesting 
feature in the study of electron transport in interferometric geometries. 
From these conductance-energy spectra we can predict that electronic 
transmission is strongly affected by the quantum interference effect 
or in other words the molecule-to-electrode interface geometry.

\subsubsection{\textcolor{dred}{Current-voltage characteristics}}

\begin{figure}[ht]
{\centering \resizebox*{8.5cm}{9cm}{\includegraphics{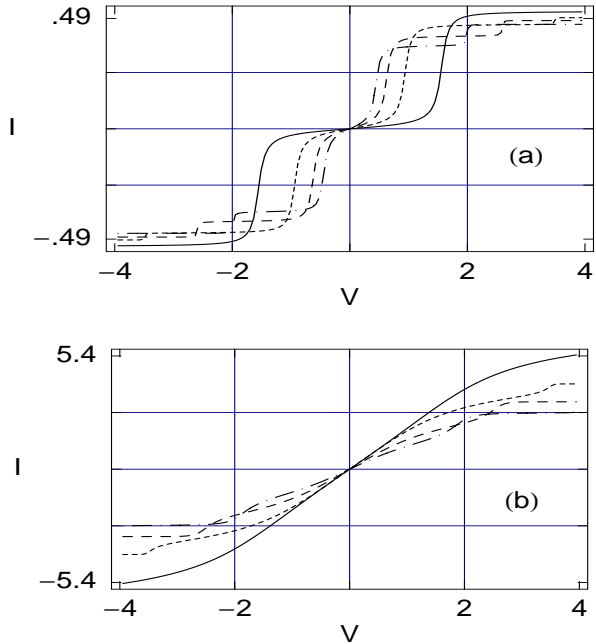}}\par}
\caption{$I$-$V$ spectra of polycyclic hydrocarbon molecules connected 
to the electrodes in the {\em trans} configuration, where (a) and (b)
correspond to the results for the weak and strong molecule-to-electrode
coupling cases, respectively. The solid, dotted, dashed and dot-dashed
lines represent the currents for the benzene, napthalene, anthracene
and tetracene molecules, respectively.}
\label{transcurr}
\end{figure}

The scenario of electron transport through these molecular wires can be 
much more clearly explained from current-voltage ($I$-$V$) spectra. 
Current through the molecular systems is computed by the integration 
procedure of transmission function $T$ (see Eq.~\ref{curr}). The behavior 
of the transmission function is similar to that of the conductance spectrum
since the relation $g=2T$ is satisfied from the Landauer conductance
formula. In Fig.~\ref{transcurr}, we plot $I$-$V$ characteristics
of the hydrocarbon molecules when they are connected in the {\em trans} 
configuration to the source and drain, where (a) and (b) correspond to the 
currents for the cases of weak- and strong-coupling limits, 
respectively. The solid, dotted, dashed and dot-dashed lines represent 
$I$-$V$ curves for the benzene, napthalene, anthracene and tetracene 
molecules, respectively. It is observed that, in the weak-coupling limit
current shows staircase-like structure with sharp steps. This is due to 
the discreteness of molecular resonances as shown by the solid curves in
Fig.~\ref{transcond}. As the voltage increases, electrochemical
potentials on the electrodes are shifted and eventually cross one of
the molecular energy levels. Accordingly, a current channel is opened
up and a jump in $I$-$V$ curve appears. The shape and height of these 
current steps depend on the width of the molecular resonances. With the 
increase of molecule-to-electrode coupling strength, current varies 
almost continuously with the applied bias voltage and achieves much 
higher values, as shown in Fig.~\ref{transcurr}(b). This continuous 
variation of the current is due to the broadening of conductance resonant 
peaks (see the dotted curves of Fig.~\ref{transcond}) in the strong 
molecule-to-electrode coupling limit.

\begin{figure}[ht]
{\centering \resizebox*{8.5cm}{9cm}{\includegraphics{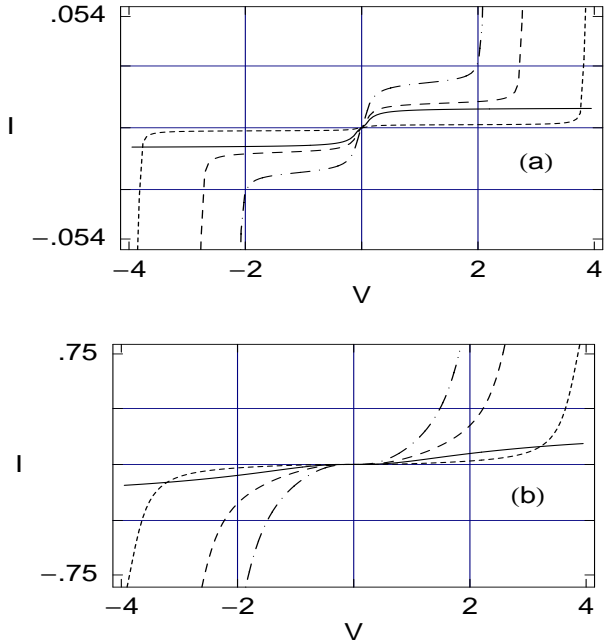}}\par}
\caption{$I$-$V$ curves of polycyclic hydrocarbon molecules connected to 
the electrodes in the {\em cis} configuration, where (a) and (b)
correspond to the results for the weak and strong molecule-to-electrode
coupling cases, respectively. The solid, dotted, dashed and dot-dashed
lines represent the identical meaning as in Fig.~\ref{transcurr}.}
\label{ciscurr}
\end{figure}

The effect of quantum interference among the electronic waves on 
molecular transport is much more clearly visible from Fig.~\ref{ciscurr}, 
where $I$-$V$ characteristics are shown for the hydrocarbon molecules 
connected to the electrodes in the {\em cis} configuration. (a) and (b) 
correspond to the currents in the two limiting cases, respectively. The 
solid, dotted, dashed and dot-dashed curves give the same meaning as in 
Fig.~\ref{transcurr}. In this configuration ({\em cis}), current amplitude 
gets reduced enormously than the case when electrodes are coupled to the
molecules in the {\em trans} configuration. This enormous change in 
current amplitude is caused solely due to the effect of quantum
interference between electronic waves passing through the molecular 
arms. Therefore, we can predict that designing a molecular device
is significantly influenced by the quantum interference effect i.e.,
molecule-to-electrode interface structure.

To summarize, in this section, we have introduced a parametric 
approach based on a simple tight-binding model to investigate electron 
transport properties of four different polycyclic hydrocarbon 
molecules sandwiched between two $1$D metallic electrodes. This approach
can be utilized to study transport behavior in any complicated molecular 
bridge system. Electron conduction through the molecular wires is 
strongly influenced by the molecule-to-electrode coupling strength and 
quantum interference effect. Our investigation provides that to design 
a molecular electronic device, in addition to the molecule itself, both
the molecular coupling and molecule-electrode interface structure are highly 
important.

\section{Designing of Classical Logic Gates}

Electronic transport in quantum confined geometries has attracted
much attention since these simple looking systems are the promising
building blocks for designing nanodevices especially in electronic
as well as spintronic engineering. The key idea of designing 
nano-electronic devices is based on the concept of quantum interference,
and it is generally preserved throughout the sample having dimension
smaller or comparable to the phase coherence length. Therefore, ring 
type conductors or two path devices are ideal candidates where the
effect of quantum interference can be exploited~\cite{san20,san21,san22}. 
In such a ring shaped geometry, quantum interference effect can be 
controlled by several ways, and most probably, the effect can be 
regulated significantly by tuning the magnetic flux, the so-called 
Aharonov-Bohm (AB) flux, that threads the ring.

In this section we will explore how a simple mesoscopic ring can be 
utilized to fabricate several classical logic gates. A single mesoscopic 
ring is used to design OR, NOT, XOR, XNOR and NAND gates, while AND and NOR
gates are fabricated with the help of two such quantum rings. For all
these logic gates, AB flux $\phi$ enclosed by a ring plays the central
role and it controls the interference condition of electronic waves 
passing through two arms of the ring. Within a non-interacting picture,
a tight-binding framework is used to describe the model and all 
calculations are done based on single particle Green's function technique.
The logical operations are analyzed by studying two-terminal conductance
as a function of energy and current as a function of applied bias 
voltage. Our numerical analysis clearly supports the logical operations
of the traditional macroscopic logic gates.

Here we describe two-input logic gates. The inputs are associated
with externally applied gate voltages through which we can tune the
strength of site energies in the atomic sites. For all logic gate
operations we fix AB flux $\phi$ at $\phi_0/2$ i.e., $0.5$ in our 
chosen unit system. The common set of values of the other parameters
used here are as follows. $\tau_S=\tau_D=0.5$ (weak-coupling limit), 
where $\tau_S$ and $\tau_D$ are the hopping integrals of a ring to the 
source and drain, respectively; $\epsilon_0=0$ and $v=4$ (parameters 
for the electrodes) and $E_F=0$.

\subsection{\textcolor{dred}{OR Gate}}

\subsubsection{\textcolor{dred}{The model}}

\begin{figure}[ht]
{\centering \resizebox*{7cm}{3.8cm}{\includegraphics{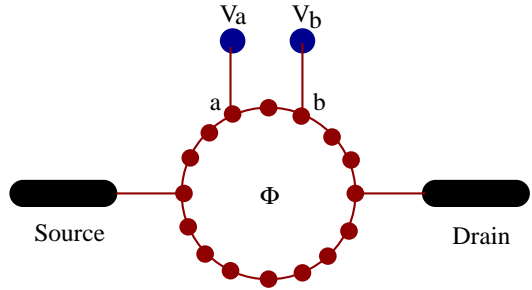}}\par}
\caption{A mesoscopic ring with total number of atomic sites $N=16$ 
(filled red circles), threaded by a magnetic flux $\phi$, is attached 
to $1$D metallic electrodes, viz, source and drain. The atoms $a$ and 
$b$ are subject to the gate voltages $V_a$ and $V_b$ respectively, 
those are variable.}
\label{or}
\end{figure}

Let us start with OR gate response. The schematic view a mesoscopic
ring that can be used as an OR gate is shown in Fig.~\ref{or}. The ring,
penetrated by an AB flux $\phi$, is symmetrically coupled (upper and lower
arms have equal number of lattice points) to two semi-infinite $1$D 
non-magnetic metallic electrodes, namely, source and drain. Two atomic
sites $a$ and $b$ in upper arm of the ring are subject to two external
gate voltages $V_a$ and $V_b$, respectively, and these are treated as 
two inputs of the OR gate. Within a non-interacting picture, the 
tight-binding Hamiltonian of the ring looks in the form,
\begin{equation}
H_{R} = \sum_i \left(\epsilon_{i0} + V_a \delta_{ia} + V_b \delta_{ib} 
\right) c_i^{\dagger} c_i 
 +  \sum_{<ij>} t \left(c_i^{\dagger} c_j e^{i\theta}+ c_j^{\dagger} 
c_i e^{-i\theta}\right).
\label{orhamil}
\end{equation}
In this Hamiltonian $\epsilon_{i0}$'s are the site energies for all the
sites $i$ except the sites $i=a$ and $b$ where the gate voltages $V_a$
and $V_b$ are applied, those are variable. These gate voltages can be
incorporated through the site energies as expressed in the above
Hamiltonian. The phase factor $\theta=2 \pi \phi/N \phi_0$ comes due to 
the flux $\phi$ threaded by the ring, where $N$ corresponds to the total 
number of atomic sites in the ring. All the other parameters have identical
meaning as described earlier. In our numerical calculations, the on-site
energy $\epsilon_{i0}$ is taken as $0$ for all the sites $i$, except the
sites $i=a$ and $b$ where the site energies are taken as $V_a$ and $V_b$,
respectively, and the nearest-neighbor hopping strength $t$ in the ring 
is set at $3$.

Quite interestingly we observe that, {\em at $\phi=\phi_0/2$ a high output
current (1) (in the logical sense) appears if one or both the inputs to the
gate are high (1), while if neither input is high (1), a low output current
(0) appears.} This phenomenon is the so-called OR gate response and here
we address it by studying conductance-energy and current-voltage 
characteristics as functions of magnetic flux and external gate 
voltages~\cite{san12}.

\subsubsection{\textcolor{dred}{Logical operation}}

\begin{figure}[ht]
{\centering \resizebox*{8cm}{7cm}{\includegraphics{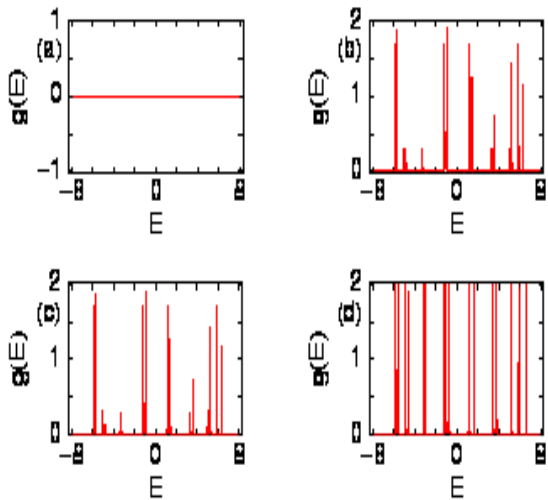}}\par}
\caption{$g$-$E$ characteristics for a mesoscopic ring with $N=16$ and 
$\phi=0.5$ in the weak-coupling limit. (a) $V_a=V_b=0$, (b) $V_a=2$ and 
$V_b=0$, (c) $V_a=0$ and $V_b=2$ and (d) $V_a=V_b=2$.}
\label{orcond}
\end{figure}

As illustrative examples, in Fig.~\ref{orcond} we display 
conductance-energy ($g$-$E$) characteristics for a mesoscopic ring
considering $N=16$ in the limit of weak ring-to-electrode coupling,
where (a), (b), (c) and (d) correspond to the results for the four
different cases of gate voltages $V_a$ and $V_b$. In the particular
case when $V_a=V_b=0$ i.e., both inputs are low ($0$), conductance
shows the value $0$ in the entire energy range (Fig.~\ref{orcond}(a)).
It indicates that electron cannot conduct from the source to drain 
across the ring. While, for the other three cases i.e., $V_a=2$ and 
$V_b=0$ (Fig.~\ref{orcond}(b)), $V_a=0$ and $V_b=2$ (Fig.~\ref{orcond}(c)) 
and $V_a=V_b=2$ (Fig.~\ref{orcond}(d)), conductance shows fine resonant 
peaks for some particular energies associated with energy eigenvalues 
of the ring. Thus, in all these three cases, electron can conduct through 
the ring. From Fig.~\ref{orcond}(d) it is observed that at resonances, 
conductance $g$ approaches the value $2$, and hence, transmission 
probability $T$ goes to unity. On the other hand, it decays from $1$ for 
the cases where any one of the two gate voltages is high and other is low 
(Figs.~\ref{orcond}(b) and (c)). Now we interpret the dependences of two 
gate voltages in these four different cases. The probability amplitude 
of getting an electron across the ring depends on the quantum interference 
of electronic waves passing through two 
arms of the ring. For symmetrically attached ring i.e., when two arms 
of the ring are identical to each other, the probability amplitude 
becomes exactly zero ($T=0$) at the typical flux $\phi=\phi_0/2$.
\begin{figure}[ht]
{\centering \resizebox*{8cm}{7cm}{\includegraphics{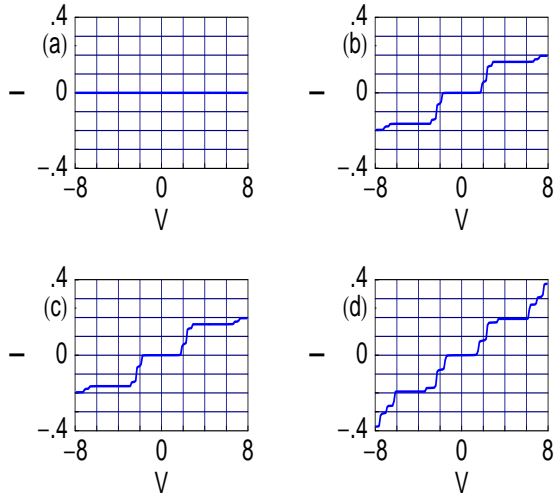}}\par}
\caption{$I$-$V$ characteristics for a mesoscopic ring with $N=16$ and 
$\phi=0.5$ in the weak-coupling limit. (a) $V_a=V_b=0$, (b) $V_a=2$ and 
$V_b=0$, (c) $V_a=0$ and $V_b=2$ and (d) $V_a=V_b=2$.}
\label{orcurr}
\end{figure}
This is due to the result of quantum interference among the waves passing 
through different arms of the ring, which can be obtained by a simple
mathematical calculation. Thus for the cases when both the two inputs
($V_a$ and $V_b$) are zero (low), the arms of the ring become identical,
and therefore, transmission probability drops to zero. On the other
hand, for the other three cases symmetry between the two arms
is broken when either the atom $a$ or $b$ or the both are subject to
the gate voltages $V_a$ and $V_b$, and therefore, the non-zero value
of transmission probability is achieved which reveals electron
conduction across the ring. The reduction of transmission probability
for the cases when any one of the two gates is high and other is low
compared to the case where both are high is also due to the
quantum interference effect. Thus we can predict that electron
conduction takes place across the ring if any one or both the inputs
to the gate are high, while if both are low, conduction is no longer 
possible. This feature clearly demonstrates the OR gate behavior. 

Following the above conductance-energy spectra now we describe the
current-voltage characteristics. In Fig.~\ref{orcurr} we display 
$I$-$V$ curves for a mesoscopic ring with $N=16$. For the case 
when both inputs are zero, the current $I$ is zero (see 
Fig.~\ref{orcurr}(a)) for the
entire bias voltage $V$. This feature is clearly visible from the
conductance spectrum, Fig.~\ref{orcond}(a), since current is
computed from integration procedure of the transmission function
\begin{table}[ht]
\begin{center}
\caption{OR gate behavior in the limit of weak-coupling. Current $I$ 
is computed at the bias voltage $6.02$.}
\label{ortable}
~\\
\begin{tabular}{|c|c|c|}
\hline \hline
Input-I ($V_a$) & Input-II ($V_b$) & Current ($I$) \\ \hline
$0$ & $0$ & $0$ \\ \hline
$2$ & $0$ & $0.164$ \\ \hline
$0$ & $2$ & $0.164$ \\ \hline
$2$ & $2$ & $0.193$ \\ \hline \hline
\end{tabular}
\end{center}
\end{table}
$T$. In the other three cases, a high output current is obtained those
are clearly presented in Figs.~\ref{orcurr}(b)-(d). The current 
exhibits staircase-like structure with fine steps as a function of 
applied bias voltage following the $g$-$E$ spectrum. In addition, it 
is also important to note that, non-zero value of the current appears
beyond a finite value of $V$, the so-called threshold voltage ($V_{th}$).
This $V_{th}$ can be controlled by tuning the size ($N$) of the ring.
From these $I$-$V$ characteristics, OR gate response is understood
very easily. 

To make it much clear, in Table~\ref{ortable}, we show a quantitative 
estimate of typical current amplitude determined at the bias voltage 
$V=6.02$. It shows that when any one of the two gates is high and other 
is low, current gets the value $0.164$, and for the case when both the 
inputs are high, it ($I$) achieves the value $0.193$. While, for the 
case when both the two inputs are low ($V_a=V_b=0$), current becomes 
exactly zero. These results clearly manifest the OR gate response in a 
mesoscopic ring.

\subsection{\textcolor{dred}{AND Gate}}

\subsubsection{\textcolor{dred}{The model}}

\begin{figure}[ht]
{\centering \resizebox*{8.4cm}{5.8cm}{\includegraphics{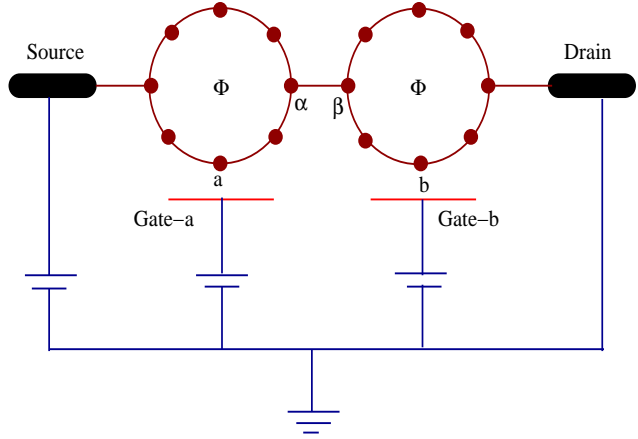}}\par}
\caption{Scheme of connections with the batteries for the operation of an 
AND gate. A double quantum ring is attached
to two semi-infinite $1$D metallic electrodes, namely, source and drain.
The gate voltages $V_a$ and $V_b$, those are variable, are applied in the
atomic sites $a$ and $b$ via the gate electrodes, gate-a and gate-b,
respectively. The source and gate voltages are applied with respect
to the drain.}
\label{and}
\end{figure}

To design an AND logic gate we use two identical quantum rings those are 
directly coupled to each other via a single bond. The schematic view of
the double quantum ring that can be used as an AND gate is presented in
Fig.~\ref{and}, where individual rings are penetrated by an AB flux $\phi$.
The double quantum ring is then attached symmetrically to two semi-infinite
$1$D metallic electrodes, namely, source and drain. Two gate electrodes, 
viz, gate-a and gate-b, are placed below the lower arms of the two rings, 
respectively, and they are ideally isolated from the rings. The atomic 
sites $a$ and $b$ in lower arms of the two rings are subject to gate 
voltages $V_a$ and $V_b$ via the gate electrodes gate-a and gate-b, 
respectively, and they are treated as two inputs of the AND gate. In 
the present scheme, we consider that the gate voltages each operate on 
the atomic sites nearest to the plates only. While, in complicated 
geometric models, the effect must be taken into account for the other 
dots, though the effect becomes too small. The actual scheme of 
connections with the batteries for the operation of the AND gate is 
clearly presented in the figure (Fig.~\ref{and}), where the source 
and gate voltages are applied with respect to the drain. The model 
quantum system is described in a similar way as prescribed
in Eq.~\ref{orhamil}. We will show that, {\em at the typical flux
$\phi=\phi_0/2$, a high output current ($1$) (in the logical sense)
appears only if both the two inputs to the gate are high ($1$), while
if neither or only one input to the gate is high ($1$), a low output
current ($0$) results.} It is the so-called AND gate response and we
investigate it by studying conductance-energy and current-voltage
characteristics~\cite{san13}.

\subsubsection{\textcolor{dred}{Logical operation}}

\begin{figure}[ht]
{\centering \resizebox*{8cm}{7cm}{\includegraphics{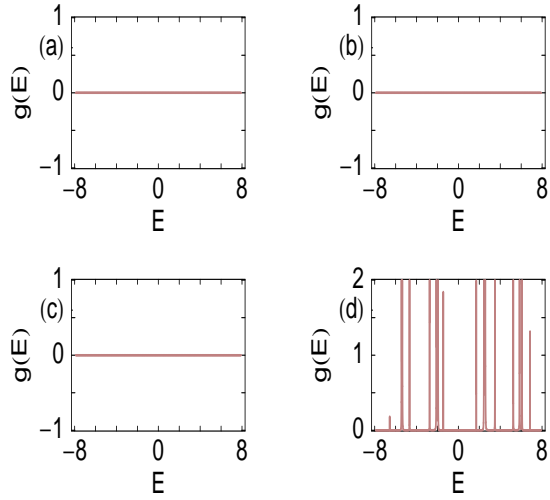}}\par}
\caption{Conductance $g$ as a function of energy $E$ for a double quantum 
ring with $M=16$ and $\phi=0.5$ in the limit of weak-coupling. 
(a) $V_a=V_b=0$, (b) $V_a=2$ and $V_b=0$, (c) $V_a=0$ and $V_b=2$ and 
(d) $V_a=V_b=2$.}
\label{andcond}
\end{figure}

In Fig.~\ref{andcond} we plot $g$-$E$ spectra for a double quantum 
ring considering $M=16$ ($M=2N$, total number of atomic sites in 
the double
quantum ring, since each ring contains $N$ atomic sites) in the limit
of weak-coupling, where (a), (b), (c) and (d) correspond to the results 
for four different choices of the gate voltages $V_a$ and $V_b$, 
respectively. When both the two inputs $V_a$ and $V_b$ are identical to 
zero i.e., both are low, conductance $g$ becomes exactly zero for the 
entire energy range (see Fig.~\ref{andcond}(a)). A similar response is 
also observed for the
other two cases where anyone of the two inputs ($V_a$ and $V_b$) to
the gate is high and other one is low. The results are shown in
Figs.~\ref{andcond}(b) and (c), respectively. Thus for all these three
cases (Figs.~\ref{andcond}(a)-(c)), the double quantum ring does not allow
to pass an electron from the source to drain. The conduction of
electron through the bridge system is allowed only when both the two
inputs to the gate are high i.e., $V_a=V_b=2$. The response is given
in Fig.~\ref{andcond}(d), and it is observed that for some particular
energies, associated with eigenenergies of the double quantum ring,
conductance exhibits fine resonant peaks. Now we try to figure out the 
dependences of gate voltages on electron transport in these four different 
cases. The probability amplitude of getting an electron from the source 
to drain across the double quantum 
ring depends on the combined effect of quantum interferences of 
electronic waves passing through upper and lower arms of the two rings. 
For a symmetrically connected ring (length of the two arms of the ring are 
identical to each other) which is threaded by a magnetic flux $\phi$, the
probability amplitude of getting an electron across the ring becomes exactly 
zero ($T=0$) at the typical flux, $\phi=\phi_0/2$. This is due to the result 
of quantum interference among the two waves in two arms of the ring, 
which can be shown through few simple mathematical steps. Thus for the 
particular case
\begin{figure}[ht]
{\centering \resizebox*{8cm}{7cm}{\includegraphics{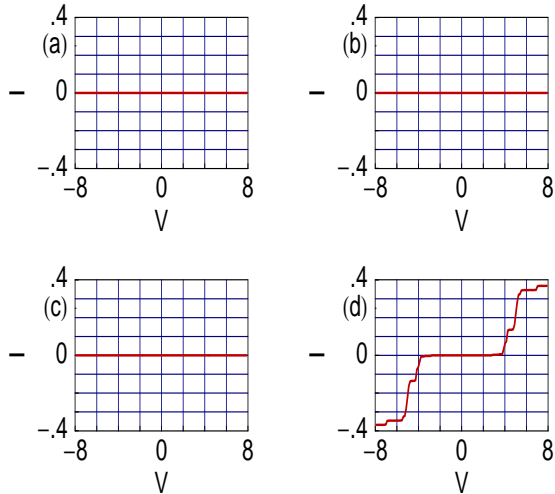}}\par}
\caption{$I$-$V$ characteristics for a double quantum ring with $M=16$ 
and $\phi=0.5$ in the weak-coupling limit. (a) $V_a=V_b=0$, (b) $V_a=2$ 
and $V_b=0$, (c) $V_a=0$ and $V_b=2$ and (d) $V_a=V_b=2$.}
\label{andcurr}
\end{figure}
when both inputs to the gate are low ($0$), the upper and lower
arms of the two rings become exactly identical, and accordingly,
transmission probability vanishes. The similar response i.e.,
vanishing transmission probability, is also achieved for the two other
cases ($V_a=2$, $V_b=0$ and $V_a=0$, $V_b=2$), where the symmetry is
broken only in one ring out of these two by applying a gate voltage
either in the site $a$ or in $b$, preserving the symmetry in the other
ring. The reason is that, when anyone of the two gates ($V_a$ and $V_b$)
is non-zero, symmetry between the upper and lower arms is broken
only in one ring which provides non-zero transmission probability across 
the ring. While, for the other ring where no gate voltage is applied, 
symmetry between the two arms becomes preserved which gives zero 
transmission probability. Accordingly, the combined effect provides 
vanishing transmission probability across the bridge, as the rings are 
coupled to each other.
The non-zero value of transmission probability is achieved only when
the symmetries of both the two rings are identically broken. This can be
done by applying gate voltages in both the sites $a$ and $b$ of the two 
rings. Thus for the particular case when both the two inputs are high i.e., 
$V_a=V_b=2$, non-zero value of the transmission probability appears. This 
feature clearly demonstrates the AND gate behavior. 

Now we go for the current-voltage characteristics to reveal AND gate 
response in a double quantum ring. As representative examples, in 
Fig.~\ref{andcurr} we plot the current $I$ as a function of applied 
bias voltage $V$ for a double quantum ring considering $M=16$ in the 
limit of weak-coupling, where (a), (b), (c) and (d) represent the results 
for four different cases of the two
\begin{table}[ht]
\begin{center}
\caption{AND gate response in the weak-coupling limit. Current $I$ is 
computed at the bias voltage $6.02$.}
\label{andtable}
~\\
\begin{tabular}{|c|c|c|}
\hline \hline
Input-I ($V_a$) & Input-II ($V_b$) & Current ($I$) \\ \hline
$0$ & $0$ & $0$ \\ \hline
$2$ & $0$ & $0$ \\ \hline
$0$ & $2$ & $0$ \\ \hline
$2$ & $2$ & $0.346$ \\ \hline \hline
\end{tabular}
\end{center}
\end{table}
gate voltages $V_a$ and $V_b$. For the cases when either both the
two inputs to the gate are low ($V_a=V_b=0$), or anyone of the two
inputs is high and other is low ($V_a=2$, $V_b=0$ or $V_a=0$, $V_b=2$),
current is exactly zero for the entire range of bias voltage.
The results are shown in Figs.~\ref{andcurr}(a)-(c), and, the vanishing
behavior of current in these three cases can be clearly understood
from the conductance spectra Figs.~\ref{andcond}(a)-(c). The non-vanishing 
current amplitude is observed only for the typical case where both the 
two inputs to the gate are high i.e., $V_a=V_b=2$. The result is shown 
in Fig.~\ref{andcurr}(d). 
From this current-voltage curve we see that the non-zero value of the 
current appears beyond a finite value of $V$, the so-called threshold 
voltage ($V_{th}$). This $V_{th}$ can be controlled by tuning the size 
($N$) of the two rings. From these results, the behavior of AND gate 
response is clearly visible. 

In the same fashion as earlier here we also make a quantitative estimate 
for the typical current amplitude as given in Table~\ref{andtable}, where 
the typical current amplitude is measured at the bias voltage $V=6.02$. 
It shows $I=0.346$ only when two inputs to the gate are high ($V_a=V_b=2$), 
while for the other three cases when either $V_a=V_b=0$ or $V_a=2$, $V_b=0$ 
or $V_a=0$, $V_b=2$, current gets the value $0$. It verifies the AND gate 
behavior in a double quantum ring.

\subsection{\textcolor{dred}{NOT Gate}}

\subsubsection{\textcolor{dred}{The model}}

\begin{figure}[ht]
{\centering \resizebox*{6.5cm}{5cm}{\includegraphics{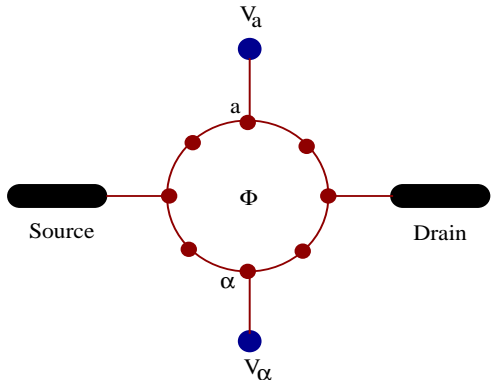}}\par}
\caption{Schematic representation for the operation of a NOT gate.
The atomic sites $a$ and $\alpha$ are subject to the voltages
$V_a$ and $V_{\alpha}$, respectively, those are variable.}
\label{not}
\end{figure}

Next we discuss NOT gate operation in a quantum ring~\cite{san14}. 
Schematic view for the operation of a NOT gate using a single mesoscopic 
ring is shown in Fig.~\ref{not}, where the ring is attached symmetrically 
to two semi-infinite $1$D metallic electrodes, viz, source and drain, 
and it is subject to an AB flux $\phi$.
A gate voltage $V_a$, taken as input voltage of the NOT gate, is applied 
to the atomic site $a$ in upper arm of the ring. While, an additional gate 
voltage $V_{\alpha}$ is applied to the site $\alpha$ in lower arm of the 
ring. Keeping $V_{\alpha}$ to a fixed value, we change $V_a$ properly to 
achieve the NOT gate operation. The model
quantum system is illustrated in the same way as given in Eq.~\ref{orhamil}. 
We will verify that, {\em at the typical flux $\phi=\phi_0/2$, a high output
current ($1$) (in the logical sense) appears if the input to the gate is
low ($0$), while a low output current ($0$) appears when the input to the
gate is high ($1$).} This phenomenon is the so-called NOT gate behavior,
and we will explore it following the same prescription as earlier.

\subsubsection{\textcolor{dred}{Logical operation}}

\begin{figure}[ht]
{\centering \resizebox*{7cm}{8.5cm}{\includegraphics{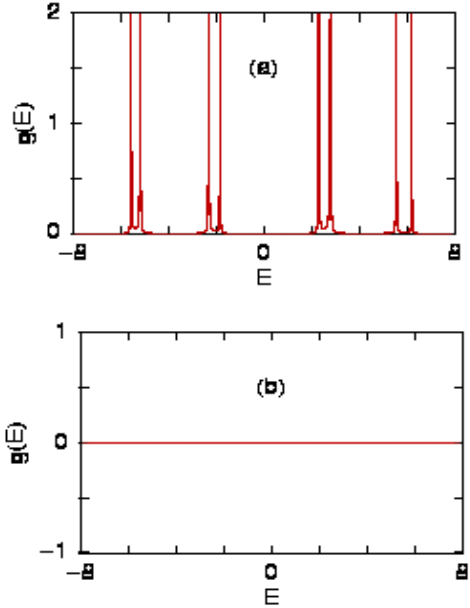}}\par}
\caption{$g$-$E$ curves in the weak-coupling limit for a mesoscopic
ring with $N=8$, $V_{\alpha}=2$ and $\phi=0.5$. (a) $V_a=0$ and
(b) $V_a=2$.}
\label{notcond}
\end{figure}

To describe NOT gate operation let us start with conductance-energy
characteristics. In Fig.~\ref{notcond} we show the variation of conductance 
as a function of injecting electron energy, in the limit of
weak-coupling, for a mesoscopic ring with $N=8$ and $V_{\alpha}=2$.
Figures~\ref{notcond}(a) and (b) correspond to the results for the
input voltages $V_a=0$ and $V_a=2$, respectively. For the particular
case when the input voltage $V_a=2$ i.e., the input is high, conductance 
$g$ vanishes (Fig.~\ref{notcond}(b)) in the complete
energy range. This indicates that conduction of electron from the 
source to drain through the ring is not possible. The situation becomes 
completely different for the case when the input to the gate is zero 
($V_a=0$). The result is shown in Fig.~\ref{notcond}(a), where conductance 
shows sharp resonant peaks for some fixed energies associated with energy 
eigenvalues of the ring. This reveals electron conduction across the ring. 
Now we focus the dependences of gate voltages on electron transport for 
two different cases of the input voltage. For the case when the input to 
the gate is
equal to $2$ i.e., $V_a=2$, the upper and lower arms of the ring become
exactly similar. This is because the potential $V_{\alpha}$ is also
set to $2$. Therefore, transmission probability drops to zero at
the typical flux $\phi_0/2$, as discussed earlier. Now if the input 
voltage $V_a$ is different from the potential applied in
the atomic site $\alpha$, then the two arms are not identical with each
other and transmission probability will not vanish. Thus, to get zero 
transmission probability when the input is high, we should tune 
$V_{\alpha}$ properly, observing the input potential and vice versa.
On the other hand, due to the breaking of symmetry between the two arms,
non-zero value of the transmission probability is achieved in the
particular case where the input voltage is zero ($V_a=0$), which reveals
electron conduction across the ring. From these results we can emphasize 
\begin{figure}[ht]
{\centering \resizebox*{7cm}{8.5cm}{\includegraphics{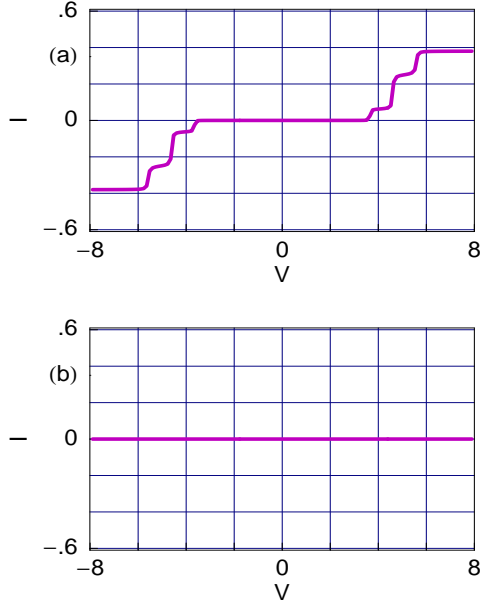}}\par}
\caption{$I$-$V$ curves in the weak-coupling limit for a mesoscopic
ring with $N=8$, $V_{\alpha}=2$ and $\phi=0.5$. (a) $V_a=0$ and
(b) $V_a=2$.}
\label{notcurr}
\end{figure}
that electron conduction takes place through the ring if the input to 
the gate is zero, while if the input is high, conduction is no longer 
possible. This aspect clearly describe the NOT gate behavior. 

To illustrate the current-voltage characteristics now we concentrate on 
the results given in Fig.~\ref{notcurr}. 
The currents are drawn as a function of applied bias voltage for a 
mesoscopic ring with $N=8$ and $V_{\alpha}=2$ in the weak-coupling limit, 
where (a) and (b) represent the results for two choices of the input 
signal $V_a$. Clearly we see that, when the input to the gate is identical
to $2$ (high), current becomes exactly zero (see Fig.~\ref{notcurr}(b)) 
for the entire bias voltage $V$. This feature is understood from the 
conductance spectrum, Fig.~\ref{notcond}(b). On the other hand, a non-zero 
value of current is obtained when the input voltage $V_a=0$, as given in 
Fig.~\ref{notcurr}(a). The current becomes non-zero beyond a threshold 
voltage $V_{th}$ which is tunable depending on the ring size and 
ring-electrode coupling strength. From these current-voltage curves it 
is clear that a high output current appears only if the input to the gate 
is low, while for high input, current doesn't appear. It justifies NOT 
gate response in the quantum ring. 

In a similar way, as we have studied earlier in other logical operations,
in Table~\ref{nottable} we make a quantitative measurement of the typical
current amplitude for the quantum ring. The current amplitude is computed
at the bias voltage $V=6.02$. It shows that, when the input to the gate 
\begin{table}[ht]
\begin{center}
\caption{NOT gate behavior in the limit of weak-coupling. Current $I$ 
is computed at the bias voltage $6.02$.}
\label{nottable}
~\\
\begin{tabular}{|c|c|}
\hline \hline
Input ($V_a$) & Current ($I$) \\ \hline
$0$ & $0.378$ \\ \hline
$2$ & $0$ \\ \hline \hline
\end{tabular}
\end{center}
\end{table}
is zero, current gets the value $0.378$. While, current becomes exactly 
zero when the input voltage $V_a=2$. Thus the NOT gate operation by using 
a quantum ring is established.

Up to now we have studied three primary logic gate operations using
one (OR and NOT) and two (AND) mesoscopic rings. In the forthcoming
sub-sections we will explore the other four combinatorial logic gate
operations using such one or two rings.

\subsection{\textcolor{dred}{NOR Gate}}

\subsubsection{\textcolor{dred}{The model}}

\begin{figure}[ht]
{\centering \resizebox*{8.5cm}{7.5cm}{\includegraphics{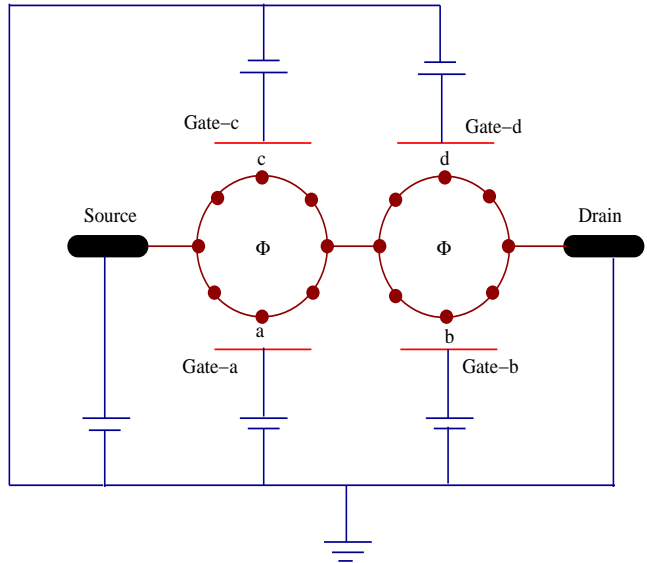}}\par}
\caption{Scheme of connections with the batteries for the operation 
of a NOR gate.}
\label{nor}
\end{figure}

We begin by referring to Fig.~\ref{nor}. A double quantum ring, where
each ring is threaded by a magnetic flux $\phi$, is sandwiched 
symmetrically between two semi-infinite one-dimensional ($1$D) metallic 
electrodes. The atomic sites $a$ and $b$ in lower arms of the two rings are
subject to gate voltages $V_a$ and $V_b$ through the gate electrodes
gate-a and gate-b, respectively. These gate voltages are variable and
treated as two inputs of the NOR gate. In a similar way we also apply
two other gate voltages $V_c$ and $V_d$, those are not varying, in the atomic
sites $c$ and $d$ in upper arms of the two rings via the gate electrodes
gate-c and gate-d, respectively. All these gate electrodes are ideally
isolated from the rings, and here we assume that the gate voltages
each operate on the atomic sites nearest to the plates only. While, in
complicated geometric models, the effect must be taken into account for
other atomic sites, though the effect becomes too small. The actual
scheme of connections with the batteries for the operation of a NOR gate 
is clearly presented in the figure (Fig.~\ref{nor}), where the
source and gate voltages are applied with respect to the drain.
The tight-binding Hamiltonian of the model quantum system is described
in a similar fashion as given in Eq.~\ref{orhamil}. Keeping $V_c$ and
$V_d$ to some specific values, we regulate $V_a$ and $V_b$ properly to 
achieve NOR gate operation. Quite nicely we establish that, {\em at the 
typical AB flux $\phi=\phi_0/2$, a high output current ($1$) (in the logical 
sense) appears if both inputs to the gate are low ($0$), while if one or 
both are high ($1$), a low output current ($0$) results.} This phenomenon 
is the so-called NOR gate response and we will illustrate it by describing 
conductance-energy and current-voltage characteristics~\cite{san15}.

\subsubsection{\textcolor{dred}{Logical operation}}

\begin{figure}[ht]
{\centering \resizebox*{8cm}{7cm}{\includegraphics{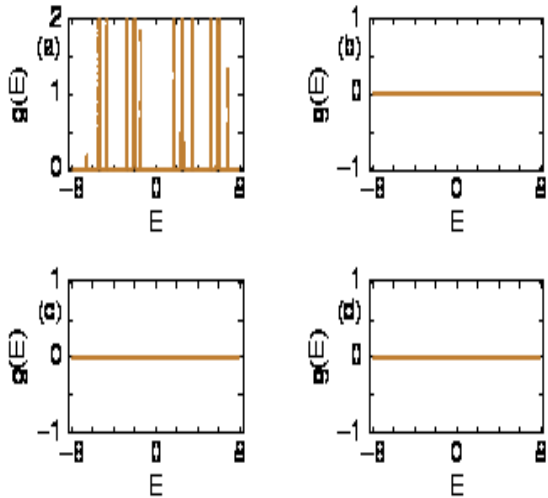}}\par}
\caption{$g$-$E$ spectra for a double quantum ring with
$M=16$, $V_c=V_d=2$ and $\phi=0.5$ in the limit of weak-coupling.
(a) $V_a=V_b=0$, (b) $V_a=2$ and $V_b=0$, (c) $V_a=0$ and $V_b=2$ and
(d) $V_a=V_b=2$.}
\label{norcond}
\end{figure}

As representative examples, in Fig.~\ref{norcond} we display the variation
of conductance $g$ as a function of injecting electron energy $E$ for 
a double quantum ring with $M=16$ ($M=2N$, total number of atomic
sites in the double quantum ring, since each ring contains $N$ atomic
sites) in the weak-coupling limit, where
(a), (b), (c) and (d) represent the results for four different cases
of the gate voltages $V_a$ and $V_b$, respectively. Quite interestingly
from these spectra we observe that, for the case when both the two inputs
$V_a$ and $V_b$ are identical to $2$ i.e., both are high, conductance
becomes exactly zero for the full range of energy $E$ (see
Fig.~\ref{norcond}(d)). An exactly similar response is also visible for
other two cases where anyone of the two inputs is high and other is
low. The results are shown in Figs.~\ref{norcond}(b) and (c), respectively.
Hence, for all these three cases (Figs.~\ref{norcond}(b)-(d)), no electron
conduction takes place from the source to drain through the double
quantum ring. The electron conduction through the bridge system is allowed
only for the typical case where both the inputs to the gates are low
i.e., $V_a=V_b=0$. The spectrum is given in Fig.~\ref{norcond}(a). It is
noticed that, for some particular energies conductance exhibits sharp
resonant peaks associated with energy levels of the double quantum ring. 
Now we try to explain the roles of the gate voltages on electron
transport in these four different cases. The probability amplitude of
getting an electron from the source to drain across the double quantum
ring depends on the combined effect of quantum interferences of the
electronic waves passing through upper and lower arms of the two rings.
For the particular case when both the two inputs to the gate are high
i.e., $V_a=V_b=2$, upper and lower arms of the two rings become exactly
identical since the gate voltages $V_c$ and $V_d$ in the upper arms are also
fixed at the value $2$. This provides the vanishing transmission probability.
If the input voltages $V_a$ and $V_b$ are different from the potential
applied in the atomic sites $c$ and $d$, then the upper and lower arms
of the two rings are no longer identical to each other and transmission
probability will not vanish. Thus, to get zero transmission probability
when the inputs are high, we should tune $V_c$ and $V_d$ properly, observing
the input potentials and vice versa. A similar behavior is also noticed
for the other two cases ($V_a=2$, $V_b=0$ and $V_a=0$, $V_b=2$), where 
symmetry is broken in only one ring out of these two by making the gate 
voltage either in the site $b$ or in $a$ to zero, maintaining the symmetry 
in other ring. The reason is that, when anyone of the two gates ($V_a$ and 
$V_b$) becomes zero, symmetry between the upper and lower arms is broken 
only in one ring which provides
\begin{figure}[ht]
{\centering \resizebox*{8cm}{7cm}{\includegraphics{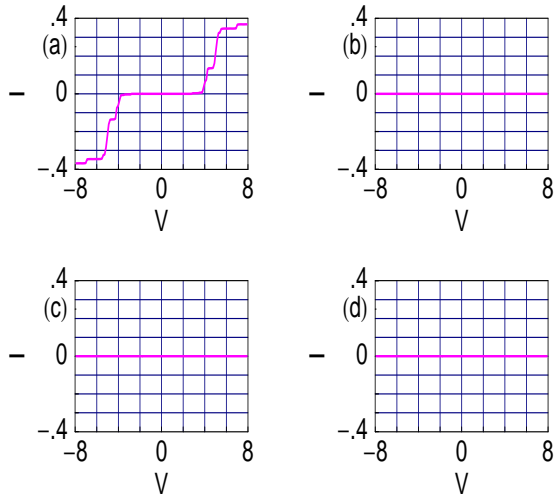}}\par}
\caption{Current $I$ as a function of applied bias voltage $V$ for a 
double quantum ring with $M=16$, $V_c=V_d=2$ and $\phi=0.5$ in the 
weak-coupling limit. (a) $V_a=V_b=0$, (b) $V_a=2$ and $V_b=0$, 
(c) $V_a=0$ and $V_b=2$ and (d) $V_a=V_b=2$.}
\label{norcurr}
\end{figure}
non-zero transmission probability across the ring. While, for the other ring
where the gate voltage is applied, symmetry between the two arms becomes
preserved which gives zero transmission probability. Accordingly, the
combined effect of these two rings provides vanishing transmission
probability across the bridge, as the rings are coupled to each other. 
The non-zero value of transmission probability appears only when 
symmetries of both the two rings are identically broken, and it is 
available for the particular case when two inputs to the gate are low i.e., 
$V_a=V_b=0$. This feature clearly demonstrates the NOR gate behavior.

To support NOR gate operation now we focus our mind on the $I$-$V$
characteristics. As illustrative examples, in Fig.~\ref{norcurr} we show 
the variation of current $I$ as a function of applied bias voltage $V$ 
for a double quantum ring with $M=16$ in the limit of weak-coupling,
where (a), (b), (c) and (d) correspond to the results for the different 
cases of two input voltages, respectively. For the cases when either 
both the two inputs to the gate are high ($V_a=V_b=2$), or anyone of the 
two inputs is high and other is low ($V_a=2$, $V_b=0$ or $V_a=0$, $V_b=2$), 
current drops exactly to zero for the whole range of bias voltage. 
The results are shown in Figs.~\ref{norcurr}(b)-(d), and, the vanishing 
behavior of current in these three different cases can be clearly 
\begin{table}[ht]
\begin{center}
\caption{NOR gate response in the limit of weak-coupling. Current
$I$ is computed at the bias voltage $6.02$.}
\label{nortable}
~\\
\begin{tabular}{|c|c|c|}
\hline \hline
Input-I ($V_a$) & Input-II ($V_b$) & Current ($I$) \\ \hline
$0$ & $0$ & $0.346$ \\ \hline
$2$ & $0$ & $0$ \\ \hline
$0$ & $2$ & $0$ \\ \hline
$2$ & $2$ & $0$ \\ \hline \hline
\end{tabular}
\end{center}
\end{table}
understood from conductance spectra given in Figs.~\ref{norcond}(b)-(d).
The finite value of current is observed only for the typical case
where both the two inputs to the gate are low i.e., $V_a=V_b=0$. The
result is shown in Fig.~\ref{norcurr}(a). At much low bias voltage,
current is almost zero and it shows a finite value beyond a threshold
voltage $V_{th}$ depending on the ring size ans ring-to-electrode
coupling strength. These features establish the NOR gate response.

For the sake of our completeness, in Table~\ref{nortable} we do a 
quantitative measurement of the typical current amplitude, determined
at $V=6.02$, for four different choices of two input signals in the 
limit of weak-coupling. Our measurement shows that current gets a finite 
value ($0.346$) only when both inputs are low ($0$). On the other hand, 
current becomes zero for all other cases i.e., if one or both inputs 
are high. Therefore, it is manifested that a double quantum ring can 
be used as a NOR gate.

\subsection{\textcolor{dred}{XOR Gate}}

\subsubsection{\textcolor{dred}{The model}}

As a follow up, now we address XOR gate response which is designed by
using a single mesoscopic ring~\cite{san16}. The ring, penetrated by 
an AB flux $\phi$,
\begin{figure}[ht]
{\centering \resizebox*{7.7cm}{6.5cm}{\includegraphics{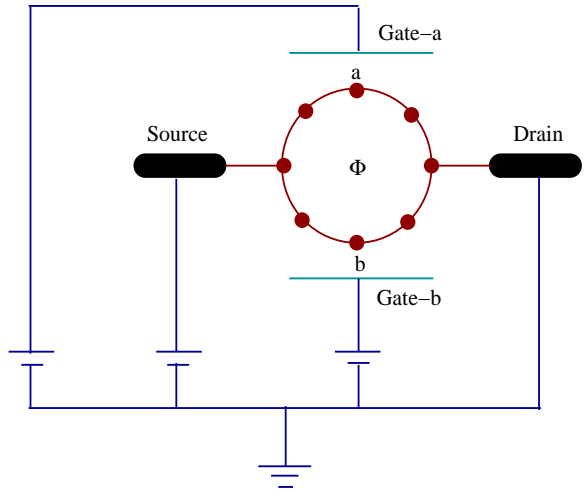}}\par}
\caption{Scheme of connections with the batteries for the operation of 
an XOR gate.}
\label{xor}
\end{figure}
is attached symmetrically to two semi-infinite $1$D non-magnetic metallic 
electrodes, namely, source and drain. The ring is placed between two
gate electrodes, viz, gate-a and gate-b. These gate electrodes are ideally
isolated from the ring and can be regarded as two parallel plates of
a capacitor. In our present scheme we assume that the gate voltages
each operate on the dots nearest to the plates only. While, in
complicated geometric models, the effect must be taken into account
for the other dots, though the effect becomes too small. The dots $a$
and $b$ in two arms of the ring are subject to gate voltages
$V_a$ and $V_b$, respectively, and these are treated as two inputs
of the XOR gate. The actual scheme of connections with the batteries
for the operation of the XOR gate is clearly presented in the
figure (Fig.~\ref{xor}), where the source and gate voltages are
applied with respect to the drain. We describe the model quantum system
through a similar kind of tight-binding Hamiltonian as presented in
Eq.~\ref{orhamil}.
Very nicely we follow that, {\em at the typical AB flux $\phi=\phi_0/2$, 
a high output current ($1$) (in the logical sense) appears if one, and 
only one, of the inputs to the gate is high ($1$), while if both inputs 
are low ($0$) or both are high ($1$), a low output current ($0$) results.} 
This is the so-called XOR gate behavior and we will emphasize it according 
to our earlier prescription.

\subsubsection{\textcolor{dred}{Logical operation}}

\begin{figure}[ht]
{\centering \resizebox*{8cm}{7cm}{\includegraphics{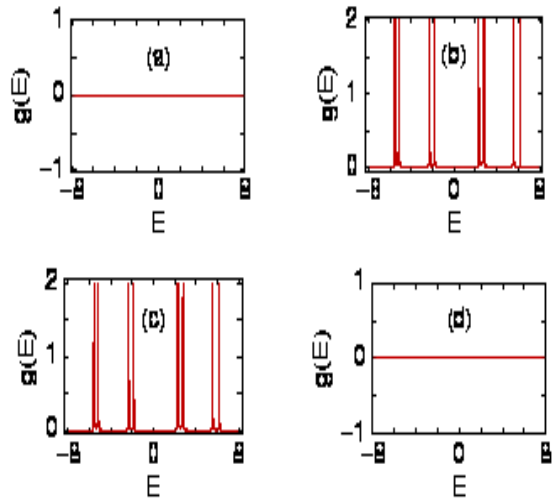}}\par}
\caption{Conductance $g$ as a function of the energy $E$ for a mesoscopic 
ring with $N=8$ and $\phi=0.5$ in the limit of weak-coupling. 
(a) $V_a=V_b=0$, (b) $V_a=2$ and $V_b=0$, (c) $V_a=0$ and $V_b=2$ and 
(d) $V_a=V_b=2$.}
\label{xorcond}
\end{figure}

Let us start with conductance-energy characteristics given in
Fig.~\ref{xorcond}. The variations of conductances are shown as a 
function of injecting electron energy $E$ for a quantum ring considering 
$N=8$ in the weak ring-to-electrode coupling limit, where four different 
figures correspond to the results for different choices of two input 
signals $V_a$ and $V_b$. When both the two inputs $V_a$ and $V_b$
are identical to zero i.e., both inputs are low, conductance $g$
becomes exactly zero (Fig.~\ref{xorcond}(a)) for all energies. This reveals
that electron cannot conduct through the ring. A similar response is also
observed when both the two inputs are high i.e., $V_a=V_b=2$, and in this
case also the ring does not allow to pass an electron from the source to
drain (Fig.~\ref{xorcond}(d)). On the other hand, for the cases
where any one of the two inputs is high and other is low i.e., either
$V_a=2$ and $V_b=0$ (Fig.~\ref{xorcond}(b)) or $V_a=0$ and $V_b=2$
(Fig.~\ref{xorcond}(c)), conductance exhibits fine resonant peaks
for some particular energies associated with energy levels of the
ring. Thus for both these two cases electron conduction takes place 
across the ring. Now we justify the dependences of gate voltages 
on electron transport for these four different cases. For the 
cases when both inputs ($V_a$ and $V_b$) are either low or high,
the upper and lower arms of the ring are exactly similar in nature,
and therefore, at $\phi=\phi_0/2$ transmission probability drops
exactly to zero. On the other hand, for the other two cases symmetry
between the arms of the ring is broken by applying a gate voltage either
in the atom $a$ or $b$, and therefore, the non-zero value of
transmission probability is achieved which reveals electron conduction
across the ring. Thus we can predict that electron conduction takes
place across the ring if one, and only one, of the inputs to the gate is
\begin{figure}[ht]
{\centering \resizebox*{8cm}{7cm}{\includegraphics{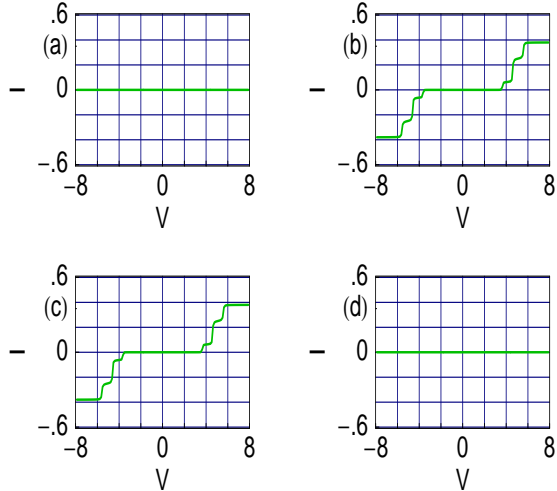}}\par}
\caption{Current $I$ as a function of bias voltage $V$ for a mesoscopic 
ring with $N=8$ and $\phi=0.5$ in the limit of weak-coupling. 
(a) $V_a=V_b=0$, (b) $V_a=2$ and $V_b=0$, (c) $V_a=0$ and $V_b=2$ and 
(d) $V_a=V_b=2$.}
\label{xorcurr}
\end{figure}
high, while if both inputs are low or both are high, conduction is
no longer possible. This phenomenon illustrates the traditional XOR 
gate behavior.

With these conductance-energy spectra (Fig.~\ref{xorcond}), now we
focus our attention on current-voltage characteristics. As illustrative 
examples, in Fig.~\ref{xorcurr} we show current-voltage characteristics 
for a mesoscopic ring with $N=8$ in the limit of weak-coupling. For the 
cases when both the two inputs are identical 
\begin{table}[ht]
\begin{center}
\caption{XOR gate behavior in the limit of weak-coupling. Current $I$ 
is computed at the bias voltage $6.02$.}
\label{xortable}
~\\
\begin{tabular}{|c|c|c|}
\hline \hline
Input-I ($V_a$) & Input-II ($V_b$) & Current ($I$) \\ \hline
$0$ & $0$ & $0$ \\ \hline
$2$ & $0$ & $0.378$ \\ \hline
$0$ & $2$ & $0.378$ \\ \hline
$2$ & $2$ & $0$ \\ \hline \hline
\end{tabular}
\end{center}
\end{table}
to each other, either low (Fig.~\ref{xorcurr}(a)) or high 
(Fig.~\ref{xorcurr}(d)), current becomes zero for the entire bias 
voltage. This behavior is clearly understood from the conductance 
spectra, Figs.~\ref{xorcond}(a) and (d). For the other two cases 
where only one of the two inputs is high and other is low, a high 
output current is obtained which are clearly described in 
Figs.~\ref{xorcurr}(b) and (c). The finite value of current appears 
when the applied bias voltage crosses a limiting value, the so-called 
threshold bias voltage $V_{th}$. Thus to get a current across the ring, 
we have to take care about the threshold voltage. These results implement 
the XOR gate response in a mesoscopic ring.

To make an end of the discussion for XOR gate response in a more compact
way in Table~\ref{xortable} we make a quantitative measurement of typical
current amplitude for the four different cases of two input signals. The
current amplitudes are computed at the bias voltage $V=6.02$. It is
observed that current becomes zero when both inputs are either low or high.
While, it (current) reaches the value $0.378$ when we set one input as high
and other as low. These studies suggest that a mesoscopic ring can
be used as a XOR gate.

\subsection{\textcolor{dred}{XNOR Gate}}

\subsubsection{\textcolor{dred}{The model}}

As a consequence now we will explore XNOR gate response and we design
this logic gate by means of a single mesoscopic ring~\cite{san17}. 
The ring, threaded by an AB flux $\phi$, is attached symmetrically 
(upper and lower arms have equal number of lattice points) to two 
semi-infinite one-dimensional ($1$D) metallic electrodes. The model 
quantum system is schematically shown in Fig.~\ref{xnor}.
Two gate voltages $V_a$ and $V_b$, taken as two inputs of the XNOR 
gate, are applied to the atomic sites $a$ and $b$, respectively, in upper 
arm of the ring. While, an additional gate voltage $V_{\alpha}$ 
is applied to the site $\alpha$ in lower arm of the ring. These 
three voltages are variable. Our quantum system is described by a
similar kind of Hamiltonian given in Eq.~\ref{orhamil}. We show that, 
{\em at the typical magnetic flux $\phi=\phi_0/2$ a high output current 
($1$) (in the logical sense) appears if both the two inputs to the
gate are the same, while if one but not both inputs are high ($1$), a low
\begin{figure}[ht]
{\centering \resizebox*{7cm}{5cm}{\includegraphics{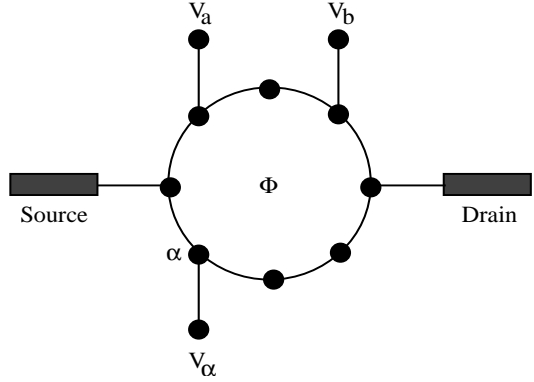}}\par}
\caption{Schematic view for the operation of a XNOR gate. The atomic
sites $a$, $b$ and $\alpha$ are subject to the voltages $V_a$, $V_b$
and $V_{\alpha}$, respectively, those are variable.}
\label{xnor}
\end{figure}
output current ($0$) results.} This logical operation is the so-called
XNOR gate behavior and we will focus it by studying conductance-energy
spectrum and current-voltage characteristics for a typical mesoscopic
ring.

\subsubsection{\textcolor{dred}{Logical operation}}

\begin{figure}[ht]
{\centering \resizebox*{8cm}{7cm}{\includegraphics{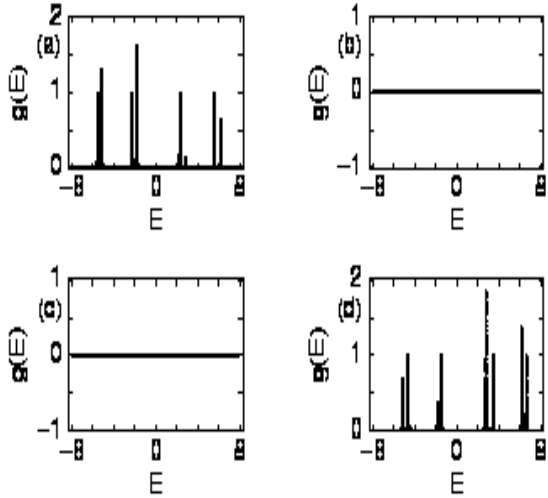}}\par}
\caption{$g$-$E$ curves in the weak-coupling limit for a mesoscopic ring
with $N=8$, $V_{\alpha}=2$ and $\phi=0.5$. (a) $V_a=V_b=0$, (b) $V_a=2$
and $V_b=0$, (c) $V_a=0$ and $V_b=2$ and (d) $V_a=V_b=2$.}
\label{xnorcond}
\end{figure}

As illustrative examples, in Fig.~\ref{xnorcond} we describe 
conductance-energy ($g$-$E$) characteristics for a mesoscopic ring
with $N=8$ and $V_{\alpha}=2$ in the limit of weak-coupling, where (a),
(b), (c) and (d) correspond to the results for different cases
of the input voltages, $V_a$ and $V_b$. For the particular cases where
anyone of the two inputs is high and other is low i.e., both inputs are 
not same, conductance becomes exactly zero (Figs.~\ref{xnorcond}(b) and 
(c)) for the whole energy range. This
predicts that for these two cases, electron cannot conduct through the
ring. The situation becomes completely different for the cases when
both the inputs to the gate are same, either low ($V_a=V_b=0$) or 
high ($V_a=V_b=2$). In these two cases, conductance exhibits fine resonant
peaks for some particular energies (Figs.~\ref{xnorcond}(a) and (d)),
which reveals electron conduction across the ring. At the resonant
energies, $g$ does not get the value $2$, and therefore, transmission
probability $T$ becomes less than unity, since the expression $g=2T$
is satisfied from the Landauer conductance formula (see Eq.~\ref{land}
with $e=h=1$). This reduction of transmission amplitude is due to the
effect of quantum interference. Now we discuss the effect of gate voltages 
on electron transport for these four different cases of the input voltages. 
For the cases when anyone of the two inputs to the gate is identical to 
$2$ and other one is $0$, the upper and lower arms of the ring become 
exactly similar. This is because the potential $V_{\alpha}$ is also set 
\begin{figure}[ht]
{\centering \resizebox*{8cm}{7cm}{\includegraphics{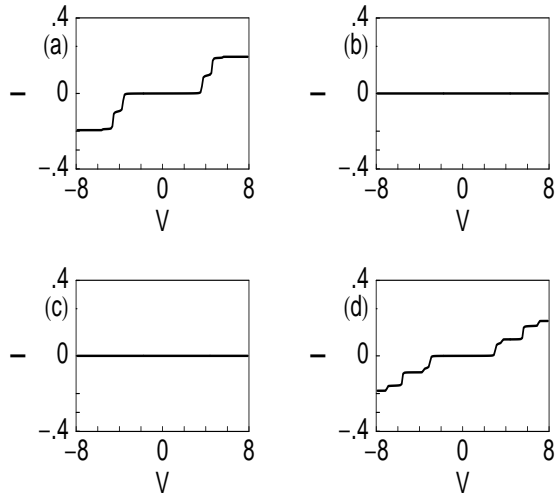}}\par}
\caption{Current $I$ as a function of the bias voltage $V$ for a
mesoscopic ring with $N=8$, $V_{\alpha}=2$ and $\phi=0.5$ in the limit
of weak-coupling. (a) $V_a=V_b=0$, (b) $V_a=2$ and $V_b=0$, (c) $V_a=0$
and $V_b=2$ and (d) $V_a=V_b=2$.}
\label{xnorcurr}
\end{figure}
to $2$. Accordingly, transmission probability $T$ drops to zero. If the 
high value ($2$) of anyone of the two gates is different from the potential 
applied in the atomic site
$\alpha$, then two arms are not identical with each other and transmission 
probability will not vanish. Thus, to get zero transmission probability 
when $V_a$ is high and $V_b$ is low and vice
versa, we should tune $V_{\alpha}$ properly, observing the input potential.
On the other hand, due to the breaking of symmetry of the two arms,
non-zero value of the transmission probability is achieved in the
particular cases when both inputs to the gate are same, which reveals 
electron conduction across the ring. From these results we can emphasize 
that electron conduction through the ring takes place if both the two 
inputs to the gate are the same (low or high), while if one but not both 
inputs are high, conduction is no longer possible. This aspect clearly 
describes the XNOR gate behavior.
 
As a continuation now we follow current-voltage characteristics to 
reveal the XNOR gate response. As representative examples, in 
Fig.~\ref{xnorcurr} we plot $I$-$V$ characteristics for a mesoscopic 
ring with $N=8$ and $V_{\alpha}=2$, in the limit of weak-coupling, where 
(a), (b), (c) and (d) represent the results for four different cases 
of the two input voltages. From these characteristics it is clearly 
observed that for the cases when one input is high and other is low, 
\begin{table}[ht]
\begin{center}
\caption{XNOR gate behavior in the limit of weak-coupling. Current $I$ 
is computed at the bias voltage $6.02$.}
\label{xnortable}
~\\
\begin{tabular}{|c|c|c|}
\hline \hline
Input-I ($V_a$) & Input-II ($V_b$) & Current ($I$) \\ \hline
$0$ & $0$ & $0.194$ \\ \hline
$2$ & $0$ & $0$ \\ \hline
$0$ & $2$ & $0$ \\ \hline
$2$ & $2$ & $0.157$ \\ \hline \hline
\end{tabular}
\end{center}
\end{table}
current is exactly zero (see Figs.~\ref{xnorcurr}(b) and (c)) for the 
entire bias voltage $V$. This phenomenon can be understood from the 
conductance spectra, Figs.~\ref{xnorcond}(b) and (c). The non-zero value 
of current appears only when both the two inputs are identical to zero 
(Fig.~\ref{xnorcurr}(a)) or high (Fig.~\ref{xnorcurr}(d)). Here also the 
current appears beyond a finite threshold voltage $V_{th}$ which is 
regulated under the variation of system size $N$ and ring-electrode 
coupling strength. These $I$-$V$ curves justify the XNOR gate response 
in a mesoscopic ring.

To be more precise, in Table~\ref{xnortable} we make a quantitative
measurement of typical current amplitude for the different choices of
two input signals in the weak ring-to-electrode coupling. The typical
current amplitude is computed at the bias voltage $V=6.02$. It is noticed
that current gets the value $0.194$ when both inputs are low ($0$), while
it becomes $0.157$ when both inputs to the gate are high ($2$). On the
other hand for all other cases, current is always zero. These results
emphasize that a mesoscopic ring can be used to design a XNOR gate.

\subsection{\textcolor{dred}{NAND Gate}}

\subsubsection{\textcolor{dred}{The model}}

At the end, we demonstrate NAND gate response and we design this logic
gate with the help of a single mesoscopic ring. A mesoscopic ring, 
threaded by a magnetic flux $\phi$, is attached symmetrically (upper and 
lower arms have equal number of lattice points) to two semi-infinite 
\begin{figure}[ht]
{\centering \resizebox*{7cm}{4.7cm}{\includegraphics{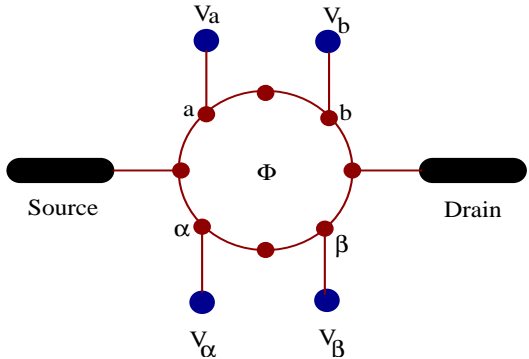}}\par}
\caption{Schematic representation for the operation of a NAND gate. The 
atoms $a$, $b$, $\alpha$ and $\beta$ are subject to the voltages $V_a$, 
$V_b$, $V_{\alpha}$ and $V_{\beta}$, respectively, those are variable.}
\label{nand}
\end{figure}
one-dimensional ($1$D) metallic electrodes. The schematic view of the 
ring that can be used to design a NAND gate is shown in Fig.~\ref{nand}. 
In upper arm of the ring two atoms $a$ and $b$ are subject to gate 
voltages $V_a$ and $V_b$, respectively, those are treated as two inputs 
of the NAND gate. On the other hand, two additional voltages $V_{\alpha}$ 
and $V_{\beta}$ are applied in the atoms $\alpha$ and $\beta$, respectively, 
in lower arm of the ring.
The model quantum system is expressed by a similar kind of tight-binding
Hamiltonian as prescribed in Eq.~\ref{orhamil}. Quite interestingly we 
notice that, {\em at the typical AB flux $\phi=\phi_0/2$ a high output 
current ($1$) (in the logical sense) appears if one or both inputs to 
the gate are low ($0$), while if both inputs to the gate are high ($1$), 
a low output current ($0$) results.} This characteristic is the so-called 
NAND gate response and we will justify it by describing conductance-energy 
and current-voltage spectra~\cite{san18}.

\subsubsection{\textcolor{dred}{Logical operation}}

\begin{figure}[ht]
{\centering \resizebox*{8cm}{7cm}{\includegraphics{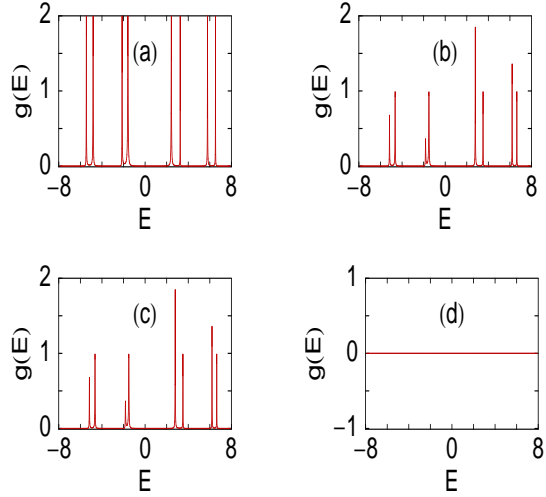}}\par}
\caption{$g$-$E$ curves in the weak-coupling limit for a
mesoscopic ring with $N=8$, $V_{\alpha}=V_{\beta}=2$ and $\phi=0.5$.
(a) $V_a=V_b=0$, (b) $V_a=2$ and $V_b=0$, (c) $V_a=0$ and $V_b=2$ and
(d) $V_a=V_b=2$.}
\label{nandcond}
\end{figure}

Let us begin with the results given in Fig.~\ref{nandcond}. Here we show 
the variation of conductance ($g$) as a function of injecting electron 
energy ($E$) in the limit of weak-coupling for a mesoscopic ring with 
$N=8$ and $V_{\alpha}=V_{\beta}=2$, where (a), (b), (c) and (d) correspond 
to the results for the different values of $V_a$ and $V_b$. When both the 
two inputs $V_a$ and $V_b$ are identical to $2$ i.e., both inputs are high, 
conductance $g$ becomes exactly zero (Fig.~\ref{nandcond}(d)) for all 
energies. This reveals that electron cannot conduct from the source to 
drain through the ring. While, for the cases where anyone or both inputs 
to the gate are zero (low), conductance gives fine resonant peaks for 
some particular energies associated with energy levels of the ring, as 
shown in Figs.~\ref{nandcond}(a), (b) and (c), respectively. Thus, in 
all these three cases, electron can conduct through the ring. From 
Fig.~\ref{nandcond}(a) it is observed that, at resonances $g$ reaches 
the value $2$ ($T=1$), but the height of resonant peaks gets down ($T<1$) 
for the cases where anyone of the two inputs is high and other is low 
(Figs.~\ref{nandcond}(b) and (c)). Now we illustrate the dependences of 
gate voltages on electron transport for these four different cases. For 
the case when both inputs to the gate are identical to $2$ i.e., 
$V_a=V_b=2$, the upper and lower arms of the ring become exactly similar.
This is due to the fact that the potentials $V_{\alpha}$ and $V_{\beta}$
are also fixed to $2$. Therefore, in this particular case transmission
probability drops to zero at $\phi=\phi_0/2$. If the two inputs $V_a$ and 
$V_b$ are different from the potentials applied in the sites $\alpha$ and 
$\beta$, then two arms are no longer identical to each other and 
transmission probability will not vanish. Hence, to get zero transmission 
probability when both inputs are high, we should tune $V_{\alpha}$ and 
$V_{\beta}$ properly, observing the input potentials and vice versa.
On the other hand, due to breaking of the symmetry of the two arms
(i.e., two arms are no longer identical to each other) in the other
\begin{figure}[ht]
{\centering \resizebox*{8cm}{7cm}{\includegraphics{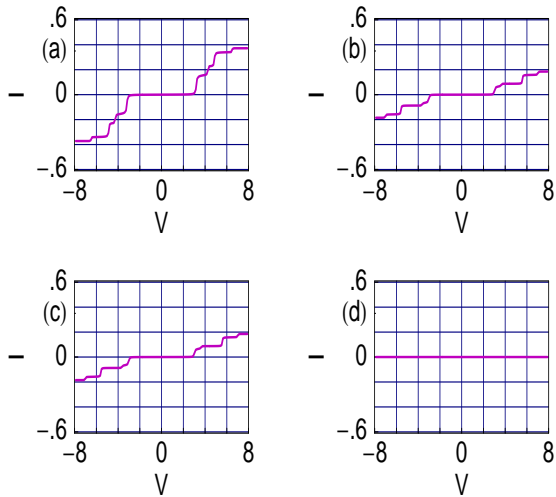}}\par}
\caption{$I$-$V$ curves in the weak-coupling limit for a
mesoscopic ring with $N=8$, $V_{\alpha}=V_{\beta}=2$ and $\phi=0.5$.
(a) $V_a=V_b=0$, (b) $V_a=2$ and $V_b=0$, (c) $V_a=0$ and $V_b=2$
and (d) $V_a=V_b=2$.}
\label{nandcurr}
\end{figure}
three cases by making anyone or both inputs to the gate are zero (low), 
the non-zero value of transmission probability is achieved which reveals 
electron conduction across the ring. The reduction of transmission 
probability from unity for the cases where any one of the
two gates is high and other is low compared to the case where both the
gates are low is solely due to the quantum interference effect. Thus
it can be emphasized that electron conduction takes place across
the ring if any one or both inputs to the gate are low, while if both 
are high, conduction is not at all possible. These results justify the 
traditional NAND gate operation.

In the same fashion, now we focus on the current-voltage characteristics.
As our illustrative purposes, in Fig.~\ref{nandcurr} we present the
variation of current $I$ in terms of applied bias voltage $V$ for a typical 
quantum ring with $N=8$ in the limit of weak-coupling, considering 
$V_{\alpha}=V_{\beta}=2$, where (a)-(d) represent the results for different 
cases of two input signals. When both inputs are high i.e, $V_a=V_b=2$ 
current becomes zero for the entire range of applied bias voltage 
(Fig.~\ref{nandcurr}(d)), following the conductance-energy spectrum given 
in Fig.~\ref{nandcond}(d). For all other three choices of two inputs, 
finite contribution in the current is available. The results are shown 
in Figs.~\ref{nandcurr}(a)-(c).
From these three current-voltage spectra, it is observed that for a fixed
bias voltage current amplitude in the typical case where both inputs are
low (Fig.~\ref{nandcurr}(a)) is much higher than the cases where one
input is high and other is low (Figs.~\ref{nandcurr}(b)-(c)). This is
clearly understood from the variations of conductance-energy spectra
studied in the above paragraph. Thus, our present current-voltage
characteristics justify the NAND gate operation in the quantum
ring very nicely.

Finally, in Table~\ref{nandtable} we present a quantitative estimate of
the typical current amplitude for four different cases of two input
signals. The typical current amplitudes are measured at the bias voltage
\begin{table}[ht]
\begin{center}
\caption{NAND gate behavior in the limit of weak-coupling. Current $I$ 
is computed at the bias voltage $6.02$.}
\label{nandtable}
~\\
\begin{tabular}{|c|c|c|}
\hline \hline
Input-I ($V_a$) & Input-II ($V_b$) & Current ($I$) \\ \hline
$0$ & $0$ & $0.339$ \\ \hline
$2$ & $0$ & $0.157$ \\ \hline
$0$ & $2$ & $0.157$ \\ \hline
$2$ & $2$ & $0$ \\ \hline \hline
\end{tabular}
\end{center}
\end{table}
$V=6.02$. It provides that current vanishes when both inputs are high
($2$). On the other hand, current gets the value $0.339$ as long as
both inputs are low ($0$) and $0.157$ when anyone of two inputs is low
and other is high. Our results support that a mesoscopic ring can be 
utilized as a NAND gate.

To summarize, in this section, we have implemented classical logic gates 
like OR, AND, NOT, NOR, XOR, XNOR and NAND using simple mesoscopic rings. 
A single ring is used to design OR, NOT, XOR, XNOR and NAND gates, while 
the rest two gates are fabricated by using two such
rings and in all the cases each ring is penetrated by an AB flux $\phi$
which plays the crucial role for whole logic gate operations. We
have used a simple tight-binding framework to describe the model, where
a ring is attached to two semi-infinite one-dimensional non-magnetic 
metallic electrodes. Based on a single particle Green's formalism all 
calculations have been done numerically which demonstrate two-terminal 
conductance and current through the system. Our theoretical analysis may 
be useful in fabricating mesoscopic or nano-scale logic gates.

Throughout the study of logic gate operations using mesoscopic rings, 
we have chosen the rings of two different sizes. In few cases we have
considered the ring with $8$ atomic sites and in few cases the number is
$16$. In our model calculations, these typical numbers ($8$ or $16$) 
are chosen only for the sake of simplicity. Though the results presented 
here change numerically with ring size ($N$), but all the basic features 
remain exactly invariant. To be more specific, it is important to note 
that, in real situation experimentally achievable rings have typical 
diameters within the range $0.4$-$0.6$ $\mu$m. In such a small ring, 
unrealistically very high magnetic fields are required to produce a quantum 
flux. To overcome this situation, Hod {\em et al.} have studied extensively 
and proposed how to construct nanometer scale devices, based on 
Aharonov-Bohm interferometry, those can be operated in moderate magnetic 
fields~\cite{hod1,hod2}.

The importance of this study is mainly concerned with (i) simplicity of 
the geometry and (ii) smallness of the size.

\section{Multi-terminal quantum transport}

Though, to date a lot of theoretical as well as experimental works
on two-terminal electron transport have been done addressing several
important issues, but a very few works are available on multi-terminal 
quantum systems~\cite{xu1,ember,let,zhong,lin,sumit1} and still it is 
an open subject to us. B\"{u}ttiker~\cite{butt2} first addressed 
theoretically the electron transport in multi-terminal quantum systems 
following the theory of Landauer two-terminal conductance formula.

In this section we investigate multi-terminal quantum transport through 
a single benzene molecule attached to metallic electrodes. A simple 
tight-binding model is used to describe the system and all calculations 
are done based on the Green's function formalism. Here we address numerical 
results which describe multi-terminal conductances, reflection probabilities 
and current-voltage characteristics. Most significantly we observe that, 
the molecular system where a benzene molecule is attached to three 
terminals can be operated as a transistor, and we call it a molecular 
transistor~\cite{san19}. This aspect can be utilized in designing 
nano-electronic circuits and our investigation may provide a basic 
framework to study electron transport in any complicated multi-terminal 
quantum system.

\subsection{\textcolor{dred}{Model and synopsis of the theoretical 
background}}

\begin{figure}[ht]
{\centering\resizebox*{5cm}{7cm}{\includegraphics{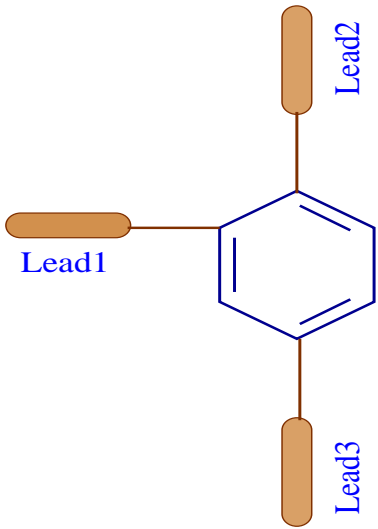}}\par}
\caption{Three-terminal quantum system where a benzene molecule is attached 
asymmetrically to three semi-infinite $1$D metallic leads, namely, lead-$1$, 
lead-$2$ and lead-$3$.}
\label{benzenethree}
\end{figure}

\begin{figure}[ht]
{\centering\resizebox*{6cm}{7.5cm}{\includegraphics{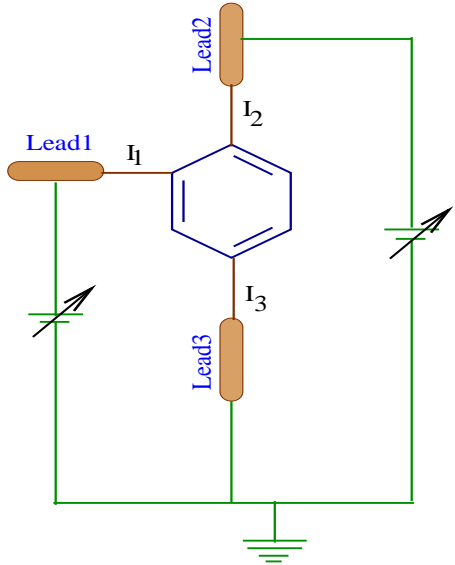}}\par}
\caption{Scheme of connections with the batteries for the operation of 
the benzene molecule as a transistor. The voltages in the lead-$1$ and 
lead-$2$ are applied with respect to lead-$3$.}
\label{circuit}
\end{figure}

Following the prescription of electron transport in two-terminal
quantum systems (as illustrated in section II) we can easily extend
our study in a three-terminal quantum system, where a benzene molecule
is attached to three semi-infinite leads, viz, lead-1, lead-2 and
lead-3. The model quantum system is shown in Fig.~\ref{benzenethree}.
These leads are coupled to the molecule asymmetrically and they are 
quite analogous to emitter, base and collector as defined in a traditional 
macroscopic transistor. The actual scheme of connections with the batteries 
for the operation of the molecule as a transistor is depicted in 
Fig.~\ref{circuit}, where the voltages in the lead-$1$ and lead-$2$ are 
applied with respect to lead-$3$. In non-interacting picture, the 
Hamiltonian of the benzene molecule can be expressed like,
\begin{equation}
H_M = \sum_i \epsilon_i c_i^{\dagger} c_i + \sum_{<ij>} 
t \left(c_i^{\dagger} c_j + c_j^{\dagger} c_i \right)
\label{benzham}
\end{equation}
where the symbols have their usual meaning. A similar kind of 
tight-binding Hamiltonian is also used to describe the side-attached leads 
which is parametrized by constant on-site potential $\epsilon_0$ and 
nearest-neighbor hopping integral $v$. We use three other parameters 
$\tau_1$, $\tau_2$ and $\tau_3$ to describe hopping integrals of the 
molecule to the lead-$1$, lead-$2$ and lead-$3$, respectively.

To calculate conductance in this three-terminal quantum system, we
use B\"{u}ttiker formalism, where all leads (current and voltage leads)
are treated in the same footing. The conductance between the terminals,
indexed by $p$ and $q$, can be written by the relation
$g_{pq}=(2e^2/h)\, T_{pq}$, where $T_{pq}$ gives the transmission
probability of an electron from the lead-$p$ to lead-$q$. Here, reflection 
probabilities are related to the transmission probabilities by the equation 
$R_{pp}+\sum_{q(\ne p)}T_{qp}=1$, which is obtained from the condition of 
current conservation~\cite{xu}.

Finally, for this three-terminal system we can write the current $I_p$ for 
the lead-$p$ as,
\begin{equation}
I_p(V)=\frac{e}{\pi \hbar} \sum_q \int \limits_{-\infty}^{\infty} 
T_{pq}(E) \left[f_p(E)-f_q(E) \right] dE
\label{equ10}
\end{equation}
Here, all the results are computed only at absolute zero temperature. 
These results are also valid even for some finite (low) temperatures, 
since the broadening of energy levels of the benzene molecule due to its 
coupling to the leads becomes much larger than that of the thermal 
broadening~\cite{datta1,datta2}. For the sake of simplicity, we take the 
unit $c=e=h=1$ in our present calculations.

\subsection{\textcolor{dred}{Numerical results and discussion}}

To illustrate the results, let us begin our discussion by mentioning the
values of different parameters used for the numerical calculations.
In the benzene molecule, on-site energy $\epsilon_i$ is fixed to $0$
for all the sites $i$ and nearest-neighbor hopping strength $t$ is
set to $3$. While, for the side-attached leads on-site energy
($\epsilon_0$) and nearest-neighbor hopping strength ($v$) are chosen 
as $0$ and $4$, respectively. The Fermi energy $E_F$ is taken as $0$. 
Throughout the study, we narrate our results for two limiting cases 
depending on the strength of coupling of the molecule to leads. 
Case I: $\tau_{1(2,3)} << t$. It is the so-called weak-coupling limit. 
For this regime we choose $\tau_1=\tau_2=\tau_3=0.5$. Case II: 
$\tau_{1(2,3)} \sim t$. This is the so-called strong-coupling limit. 
In this particular limit, we set the values of the parameters as
$\tau_1=\tau_2=\tau_3=2.5$.

\subsubsection{\textcolor{dred}{Conductance-energy characteristics}}

\begin{figure}[ht]
{\centering \resizebox*{10cm}{11cm}{\includegraphics{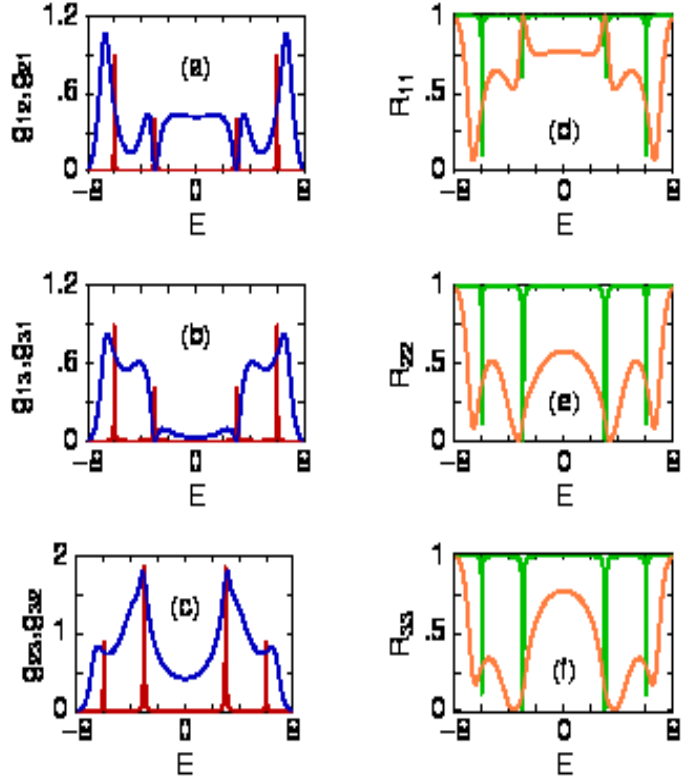}}\par}
\caption{Three-terminal conductance $g_{pq}$ and
reflection probability $R_{pp}$ as a function of energy $E$ of the
benzene molecule. The red and green curves correspond to the results
for the weak-coupling limit, while the blue and reddish yellow lines
represent the results for the limit of strong-coupling. Conductance
is measured in unit of $e^2/h$, while the energy is measured in unit
of $t$.}
\label{threecond}
\end{figure}

In three-terminal molecular system, several anomalous features are 
observed in conductance-energy spectra as well as in the variation 
of reflection probability with energy $E$. As representative 
examples, in Fig.~\ref{threecond} we present the results, where first 
column gives the variation of conductance $g_{pq}$ and second column 
represents the nature of reflection probability $R_{pp}$.
From the conductance spectra it is observed that conductances
exhibit fine resonant peaks (red curves) for some particular energies
in the limit of weak-coupling, while they get broadened (blue curves)
as long as coupling strength is enhanced to the strong-coupling
limit. The explanation for the broadening of resonant peaks is
exactly similar as described earlier in the case of two-terminal
molecular system. A similar effect of molecular coupling to the side
attached leads is also noticed in the variation of reflection
probability versus energy spectra (right column of Fig.~\ref{threecond}). 
Since in this three-terminal molecular system leads are connected 
asymmetrically to the molecule i.e., path length between the leads are 
different from each other, all conductance spectra are different in nature. 
It is also observed that the heights of different conductance peaks are 
not identical. This is solely due to the effect of quantum interference 
among different arms of the molecular ring. Now, in the variation of
reflection probabilities, we also get the complex structure like as
conductance spectra. For this three-terminal system since reflection 
probability is not related to the transmission probability simply as in 
the case of a two-terminal system, it is not necessarily true that 
$R_{pp}$ shows picks or dips where $g_{pq}$ has dips or picks. It depends 
on the combined effect of $T_{pq}$'s.

\subsubsection{\textcolor{dred}{Current-voltage characteristics: 
Transistor operation}}

Finally, we describe current-voltage characteristics for this
three-terminal molecular system and try to illustrate how it can be
operated as a transistor.

The current $I_p$ passing through any lead-$p$ is obtained by
integration procedure of the transmission function $T_{pq}$ (see
\begin{figure}[ht]
{\centering \resizebox*{8cm}{4.5cm}{\includegraphics{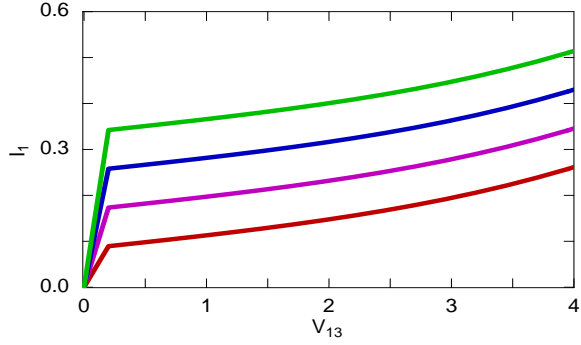}}\par}
\caption{Current $I_1$ as a function of $V_{13}$
($=V_1-V_3$) for constant $V_{12}$ ($=V_1-V_2$) for the three-terminal
molecular system in the limit of strong-coupling. The red, magenta, blue
and green curves correspond to $V_{12}=0.2$, $0.4$, $0.6$ and $0.8$,
respectively. Current is measured in unit of $et/h$, and the bias
voltage is measured in unit of $t/e$.}
\label{threecurr1}
\end{figure}
\begin{figure}[ht]
{\centering \resizebox*{8cm}{4.5cm}{\includegraphics{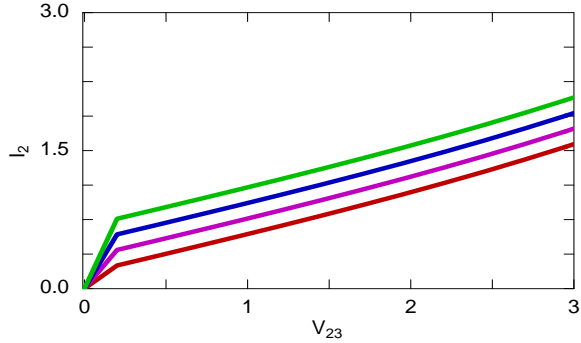}}\par}
\caption{Current $I_2$ as a function of $V_{23}$
($=V_2-V_3$) for constant $V_{12}$ for the three-terminal molecular
system in the limit of strong-coupling. The red, magenta, blue and green
curves correspond to $V_{12}=0.4$, $0.8$, $1.2$ and $1.6$, respectively.
Current is measured in unit of $et/h$, and the bias voltage is measured
in unit of $t/e$.}
\label{threecurr2}
\end{figure}
Eq.~(\ref{equ10})), where individual contributions from  the other two 
leads have to be taken into account. To be more precise, we can write 
the current expression for the three-terminal molecular device 
where one of the terminals serves as a voltage as well as a current 
probe~\cite{datta1} in the form $I_p=\sum \limits_q g_{pq} 
\left(V_p-V_q\right) \equiv \sum \limits_q g_{pq} V_{pq}$, where 
$V_{pq}=\left(V_p-V_q\right)$ is the voltage difference between the 
lead-$p$ and lead-$q$.

In Fig.~\ref{threecurr1}, we plot current $I_1$ in the lead-$1$ as
a function of $V_{13}$ for constant $V_{12}$ in the limit of strong
molecular coupling. The red, magenta, blue and green curves correspond 
to the currents for $V_{12}=0.2$, $0.4$, $0.6$ and $0.6$, respectively. 
From the results it is observed that, for a constant voltage difference 
between the lead-$1$ and lead-$2$, current $I_1$ initially rises to a 
large value when $V_{13}$ starts to increase from zero value, and after 
that, it ($I_1$) increases very slowly with the rise of $V_{13}$ and 
eventually saturates.
On the other hand, for a constant lead-$1$ to lead-$3$ voltage difference,
current $I_1$ increases gradually as we increase $V_{12}$, which is
clearly described from the four different curves in Fig.~\ref{threecurr1}.
Quite similar behavior is also observed in the variation of current
$I_2$ as a function of $V_{23}$ for constant $V_{12}$. The results are
shown in Fig.~\ref{threecurr2}, where currents are calculated for
the strong-coupling limit.
The red, magenta, blue and green lines represent the currents for
$V_{12}=0.4$, $0.8$, $1.2$ and $1.6$, respectively. Comparing the
\begin{figure}[ht]
{\centering \resizebox*{8cm}{4.5cm}{\includegraphics{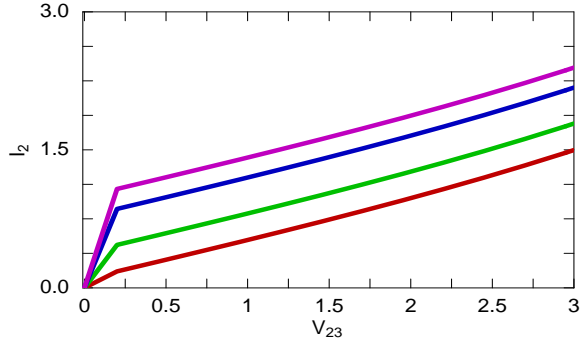}}\par}
\caption{Current $I_2$ as a function of $V_{23}$ for
constant $I_1$ for the three-terminal molecular system in the limit of
strong-coupling. The red, green, blue and magenta curves correspond to
$I_1=0.09$, $0.36$, $0.73$ and $0.92$, respectively. Current is measured
in unit of $et/h$, and the bias voltage is measured in unit of $t/e$.}
\label{threecurr3}
\end{figure}
results plotted in Figs.~\ref{threecurr1} and \ref{threecurr2}, it
is clearly observed that current in the lead-$2$ is much higher
than the current available in the lead-$1$ for the entire voltage
range. This is solely due to the effect of quantum interference
among the electronic waves passing through different arms of the
molecular ring, and, we can manifest that for a fixed molecular coupling, 
current amplitude significantly depends on the positions of different 
leads.

At the end, we illustrate the results plotted in Fig.~\ref{threecurr3},
where the variation of current $I_2$ is shown as a function of
$V_{23}$ for constant current $I_1$. The currents ($I_2$) are
calculated for the strong-coupling limit, where
the red, green, blue and magenta curves correspond to fixed $I_1=0.09$,
$0.36$, $0.73$ and $0.92$, respectively. For a constant $V_{23}$,
current through the lead-$2$ increases gradually as we increase the
current $I_1$ which is clearly visible from the four different curves
in this figure. These current-voltage characteristics are quite analogous
to a macroscopic transistor. Thus, in short, we can predict that this
three-terminal molecular system can be operated as a transistor and
we may call it a molecular transistor. Like a conventional macroscopic
transistor, the three different terminals of the molecular transistor
can be treated as emitter, base and collector. Here, the important point
is that, since all these three terminals are equivalent to each other,
any one of them can be considered as an emitter or base or collector.
Not only that, for this molecular transistor as there is only one type
of charge carrier, which is electron, conventional biasing method is not 
required. These features provide several key ideas which motivate us to 
develop a molecular transistor rather than the traditional one.

All the above current-voltage characteristics for the three-terminal
quantum system are studied only for the limit of strong molecular
coupling. Exactly similar features, except the current amplitude,
are also observed for the case of weak-coupling limit, and in obvious 
reason here we do not plot these results once again.

To summarize, in this section, we have explored electron transport 
through a benzene molecule attached to metallic electrodes. The molecular
system is described by a simple tight-binding Hamiltonian and all 
calculations are done through Green's function approach. We have
numerically calculated the multi-terminal conductances, reflection
probabilities and current-voltage characteristics. Very interestingly 
we have seen that the three-terminal
benzene molecule can be operated as an electronic transistor, and we
call it as a molecular transistor. These three terminals are analogous
to emitter, base and collector as defined in traditional transistor.
All these features of electron transport may be utilized in fabricating
nano-electronic devices and our detailed investigation can provide a
basic theoretical framework to characterize electron transport in any
complicated multi-terminal quantum system.

Here we have made some minor approximations in considering the site 
energies of carbon-type sites and the nearest-neighbor hopping strength 
between these sites in molecular ring. The absolute value of 
nearest-neighbor hopping strength ($t$) alternates between the two values, 
those are respectively taken as $2.5$ and $2.85$~\cite{low}. Here we have 
approximated it ($t$) to the absolute value $3$. With this assumption, the 
characteristic
features are not changed at all. On the other hand, the absolute value
of on-site energy ($\epsilon_i$) corresponding to carbon-type sites
in this molecule is identical to $6.6$~\cite{low}. But here we have set
it to zero, only for the sake of simplicity, since the constant value
of on-site energy simply gives the shift of energy eigenvalues and
hence transmission spectra, keeping the current-voltage characteristics 
unchanged. The main motivation of considering a benzene molecule to reveal 
the transistor operation is that molecular systems are ideal candidates 
for future development of nano-electronic devices. Here we have done a 
model calculation
considering a benzene molecule and presented the results for these set
of parameter values. Instead of a benzene molecule one can also take a
mesoscopic ring and perform all the calculations for other set of
parameter values. In that situation, only the results presented here
change numerically with the ring size, but all the basic features
remain exactly invariant.

\section{Summary and Conclusions}

In this review article we have demonstrated some important issues on 
electron transport through some model quantum systems. With a brief 
introduction of several fundamental issues in mesoscopic region, we 
have investigated electron transport through four different polycyclic
hydrocarbon molecules, namely, benzene, napthalene, anthracene and
tetracene. A tight-binding framework is used to describe the model
quantum systems and all calculations are performed using Green's 
function technique. Electron conduction through these molecular
wires is significantly influenced by the molecule-to-electrode
coupling strength and quantum interference of electronic waves passing
through different arms of the molecular ring. Our numerical results
predict that to design a molecular electronic device, not only the
molecule, both the molecular coupling to side attached electrodes
and molecule-electrode interface structure are highly important.
Our model calculations provide a physical insight to the behavior of
electron conduction through molecular bridge systems.

In the next part we have illustrated the possibilities of designing
classical logic gates using simple mesoscopic rings. First we have
studied three primary logic gate operations using one (OR and NOT)
and two (AND) quantum rings. Later, we have established other four
combinatorial logic gate operations (NOR, XOR, XNOR and NAND) using 
such one or two rings. The key controlling parameter for all logic 
gate operations is the magnetic flux $\phi$ threaded by the ring. The 
logical operations have been described in terms of conductance-energy 
and current-voltage characteristics. Our analysis may be useful in 
fabricating meso-scale or nano-scale logic gates.

In the last part of this review article, we have discussed about 
multi-terminal transport problem. In a multi-terminal quantum system 
electron transport cannot be addressed by using Landauer approach. 
B\"{u}ttiker first
resolved this issue considering Landauer two-terminal transport formula 
and the technique has been named as Landauer-B\"{u}ttiker approach.
Using this methodology we have studied multi-terminal quantum transport
through a single benzene molecule. Our numerical results describe 
multi-terminal conductances, reflection probabilities and current-voltage
characteristics. Quite interestingly we have shown that the three-terminal
benzene molecule can be operated as an electronic transistor. These
three terminals are analogous to the emitter, base and collector as defined
in conventional transistor. Our presented results may provide a basic
theoretical framework to address electron transport in any multi-terminal
quantum system.

\end{document}